\pdfoutput=1

\documentclass[12pt,usenames,dvipsnames]{article}

\usepackage{latexsym}
\usepackage{amsfonts, amsmath, amsthm, amssymb, mathtools}
\usepackage{graphicx}
\usepackage{indentfirst}
\usepackage{bbm}
\usepackage{verbatim}
\usepackage{mathrsfs}
\usepackage{hyperref}
\usepackage{dsfont}
\usepackage{cite}
\usepackage{xcolor}
\usepackage{enumerate}
\usepackage{cleveref}
\usepackage{physics}
\usepackage{cases}
\usepackage{appendix}

\parskip=\medskipamount

\newcommand {\cL}{{\cal L}}

\newcommand {\cN}{{\cal N}}

\def\a{\alpha}

\def\b{\beta}
\def\c{\chi}
\def\d{\delta}

\def\f{\phi}

\def\j{\psi}

\def\l{\lambda}
\def\m{\mu}
\def\n{\nu}
\def\o{\omega}

\def\r{\rho}
\def\s{\sigma}

\def\x{\xi}

\def\rd{{\rm d}}
\def\ri{{\rm i}}
\def\re{{\rm e}}

\def\rc{{\rm c}}
\def\rh{{\rm h}}

\newcommand{\ve}{\varepsilon}                            

\newcommand{\pa}{\partial}                           
\newcommand{\hf}{\frac12}

\newcommand{\vf}{\varphi}

\renewcommand{\theequation}{\thesection.\arabic{equation}}
\newcommand{\be}{\begin{equation}}
\newcommand{\ee}{\end{equation}}
\newcommand{\bea}{\begin{eqnarray}}
\newcommand{\eea}{\end{eqnarray}}
\newcommand{\non}{\nonumber}
\def\double #1{#1{\hbox{\kern-2pt $#1$}}}

\makeatletter
\renewcommand*{\p@subsection}{\thesection~}
\renewcommand{\thesection}{\Roman{section}} 
\renewcommand{\thesubsection}{\Alph{subsection}} 
\renewcommand\section{\@startsection {section}{1}{\z@}%
                                   {-3.5ex \@plus -1ex \@minus -.2ex}%
                                   {2.3ex \@plus.2ex}%
                                   {\centering\normalfont\bfseries\MakeUppercase}}
\renewcommand\subsection{\@startsection{subsection}{2}{\z@}%
                                     {-3.25ex\@plus -1ex \@minus -.2ex}%
                                     {1.5ex \@plus .2ex}%
                                     {\centering\normalfont\bfseries}}
\makeatother

\newif\ifdtup

\newcommand{\bsubeq}{\begin{subequations}}
\newcommand{\esubeq}{\end{subequations}}

\numberwithin{equation}{section}

\renewcommand{\theequation}{\arabic{section}.\arabic{equation}}

\begin{document}
\begin{titlepage}
\begin{flushright}
May, 2024 \\
Revised version: October, 2024
\end{flushright}
\vspace{5mm}

\begin{center}
{\Large \bf 
Equivalence of gauge-invariant models for massive integer-spin fields
}
\end{center}

\renewcommand{\thefootnote}{\fnsymbol{footnote}}

\begin{center}

{\bf Arcadia John Fegebank\footnote{arcadia.fegebank@research.uwa.edu.au} and Sergei M. Kuzenko\footnote[2]{sergei.kuzenko@uwa.edu.au}} \\
\vspace{5mm}

\footnotesize{
{\it Department of Physics M013, The University of Western Australia\\
35 Stirling Highway, Perth W.A. 6009, Australia}}  
~\\
\vspace{2mm}

\end{center}

\begin{abstract}
\baselineskip=14pt
There are several approaches to formulate gauge-invariant models for massive integer-spin fields in $d$ dimensions including the following: (i) in terms of symmetric tensor fields $\phi_{\mu_1 \dots \mu_k} $, with $k = s, s-1, \dots , 0$, restricted to be double traceless for $k\geq 4$; and (ii) in terms of a quartet of {\it traceful} symmetric tensor fields $\j_{\mu_1 \dots \mu_k} $, of rank $k=s,s-1,s-2, s-3$. We demonstrate that these formulations in Minkowski space ${\mathbb M}^d$ are equivalent to the gauge-invariant theory for a massive integer-spin field proposed in 1989 by Pashnev. We also make use of the Klishevich-Zinoviev theory in ${\mathbb M}^d$ to derive a unique generalization of the Singh-Hagen model for a massive integer-spin field in $d>4 $ dimensions. 
\end{abstract}
\vspace{5mm}

\vfill
\renewcommand{\thefootnote}{\arabic{footnote}}
\vfill
\end{titlepage}

\newpage
\renewcommand{\thefootnote}{\arabic{footnote}}
\setcounter{footnote}{0}

\tableofcontents{}
\vspace{1cm}
\bigskip\hrule

\allowdisplaybreaks


\section{Introduction} \label{Intro} 

Long ago \cite{Dirac,Fierz,FierzPauli}, the equations describing an on-shell massive field of arbitrary spin $s$ in four dimensions were derived. In the integer-spin case, the equations are 
\begin{subequations}\label{FPequations}
\bea
\pa^\n \vf_{\n \m_1 \dots \m_{s-1}} =0~, \qquad (\Box -m^2 ) \vf_{\m_1 \dots \m_s} =0~, \label{Fierz Pauli}
\eea 
where the dynamical field $\vf_{\m_1 \dots \m_s} $ is symmetric and traceless, 
\bea
 \vf_{(\m_1 \dots \m_s)}= \vf_{\m_1 \dots \m_s} 
~, \qquad \eta^{\n \r} \vf_{\n \r \m_1 \dots \m_{s-2}} =0~.
\eea 
\end{subequations}
In order to realize these equations as Euler-Lagrange equations in a Lagrangian field theory, it was pointed out 
that certain auxiliary fields are required for $s>1$ \cite{FierzPauli}.
A correct set of auxiliary fields and an action principle were found by Singh and Hagen in the bosonic \cite {Singh:1974}
and the fermionic \cite{SinghHagen2} cases. One may think of the integer-spin model of \cite {Singh:1974} as a higher-spin generalization of  the massive spin-two model proposed by Fierz and Pauli
\cite{FierzPauli}, 
\bea
\mathcal{L} &= &\frac{1}{2}\varphi^{\mu \n}(\square-m^2)\varphi_{\mu \n}+\partial_\nu\varphi^{\nu\mu}\partial^\lambda\varphi_{\lambda\mu} 
   +\frac{1}{3}\vf \Big\{ 2 \partial_\mu\partial_\nu\varphi^{\mu\nu}
   - \big(\square-2 m^2\big)\varphi \Big\}~.
\eea
Considering a massless limit of the  Singh-Hagen models allowed Fronsdal and Fang to derive gauge-invariant formulations 
for massless higher-spin bosonic \cite{Fronsdal:1978}  and fermionic 
\cite{FF} fields (see also \cite{Curtright:1979uz}).

It is well known that the massive spin-one model (or the Proca theory)
\bea
\cL = - \frac 14 F^{\m \n} F_{\m\n } - \hf m^2 A^\m A_\m ~, \qquad F_{\m\n} = \pa_\m A_\n - \pa_\n A_\m ~,
\label{1.3}
\eea
has a gauge-invariant Stueckelberg  reformulation
\bea
\widetilde{\cL} = - \frac 14 F^{\m \n} F_{\m\n } - \hf \pa^\m \vf \pa_\m \vf - \hf m^2 A^\m A_\m +m A^\m \pa_\m \vf~,
\label{1.4}
\eea
which is obtained from \eqref{1.3} by replacing $A_\m \to A_\m - m^{-1} \pa_\m \vf$. By construction, the gauge invariance of $\widetilde{\cL}$ is
\bea
\d A_\m = \pa_\m \x~, \qquad \d \vf = m \x~,\label{Stueckelberg Gauge Transformations}
\eea
with the gauge parameter $\x $ being arbitrary. This local symmetry allows one to choose the gauge condition $\vf =0$, and then $\widetilde{\cL}$ turns into \eqref{1.3}. 

The Singh-Hagen model \cite {Singh:1974} is a non-gauge theory. For various reasons, it is of interest to have its gauge-invariant reformulation being similar to that for the massive spin-one model just discussed. For the spin values $s=2$ and $s=3$, such reformulations were derived forty years ago by Zinoviev \cite{Zinovev1983}. In 1997, his results were extended by Klishevich and Zinoviev  to the case of an arbitrary integer spin in four-dimensional Minkowski space in \cite{Klishevich:1997pd}.\footnote{It was also shown by the authors of \cite{Klishevich:1997pd} that the Singh-Hagen theory is obtained from their formulation by appropriately fixing the gauge freedom.}
Finally, a gauge-invariant formulation for massive particles of arbitrary integer spin 
was constructed by Zinoviev \cite{Zinoviev:2001}
in a $d$-dimensional (anti-)de Sitter space (A)dS$_d$.
The construction of \cite{Zinoviev:2001} inspired Metsaev to propose a gauge-invariant formulation for massive totally symmetric fermionic fields in 
(A)dS$_d$ \cite{Metsaev}.\footnote{In Minkowski space ${\mathbb M}^d$, gauge-invariant formulations for massive fermionic higher-spin  fields were proposed earlier in \cite{Klishevich:1998yt, Buchbinder:2006nu}.}
The gauge-invariant formulations of \cite{Zinoviev:2001,Metsaev}
  have been used for various applications and generalizations, including the frame-like gauge-invariant formulation for massive high spin fields \cite{Zinoviev:2008ze} and Lagrangian descriptions of massive $\cN=1$ supermultiplets with arbitrary superspin \cite{Zinoviev:2007js, Buchbinder:2019dof}.

The gauge fields in the massive spin-$s$ Klishevich-Zinoviev (KZ) theory are symmetric rank-$k$
 tensor fields $\f^{(k)} $, with $k = s, s-1, \dots , 0$, restricted to be 
 double traceless for $k\geq 4$.
The corresponding gauge parameters are symmetric rank-$k$ tensors $\x^{(k)}$, with $k = s-1, \dots , 0$, restricted to be traceless for $k\geq 2$. It was pointed out in \cite{Klishevich:1997pd} that all the gauge fields may be packaged into a quartet of  {\it traceful} symmetric tensor fields $  \psi^{(k)} $ of rank
$k=s,s-1,s-2, s-3$. Similarly, all the gauge parameters may be packaged into two 
{\it traceful} symmetric  parameters $\ve^{(s-1)}$  and 
$\ve^{(s-2)}$. It was further mentioned that the latter gauge fields and the gauge parameters have the same tensor types as those originating in the gauge-invariant model for the massive spin-$s$ field proposed by Pashnev in 1989 \cite{Pashnev:1989}.\footnote{This paper was submitted to the journal {\it Theoretical and Mathematical Physics} in August 1987. It was published in March 1989.} 
No further comparison between the KZ and Pashnev theories was attempted in  \cite{Klishevich:1997pd}, however the main construction of \cite{Pashnev:1989} has been used in two follow-up papers by Klishevich
\cite{Klishevich:1998ng, Klishevich:1998yt}.
It will be shown in Section \ref{Pashnev} that these theories are equivalent. 

Pashnev's gauge-invariant formulation for the massive integer spin-$s$ field in ${\mathbb M}^d$ was obtained by dimensional reduction, ${\mathbb M}^{d+1} \to {\mathbb M}^d \times S^1$,
of Fronsdal's action \cite{Fronsdal:1978} (see also \cite{Curtright:1979uz}) for the massless spin-$s$ field in $(d+1)$ dimensions. The idea to generate massive higher-spin models via dimensional reduction of massless gauge fields had been advocated earlier by Aragone et al.  \cite{Aragone:1987dtt}, although they explicitly worked out only a few low spin cases. Dimensional reduction of Fronsdal's bosonic \cite{Fronsdal:1978, Curtright:1979uz} and fermionic \cite{FF} actions has been used to generate gauge-invariant higher-spin models 
in many publications 
\cite{Rindani:1988gb, Rindani:1989ym, Buchbinder:2008ss, Asano:2019smc, Lindwasser}.\footnote{The procedure of dimensional reduction of massless mixed symmetry fields in ${\mathbb M}^{d+1}$ to massive ones in ${\mathbb M}^d $ was developed in \cite{Chekmenev:2019ayr} building on the construction given in \cite{Alkalaev:2017hvj}.}
In the integer-spin case, the models derived in \cite{Buchbinder:2008ss, Asano:2019smc, Lindwasser} may be seen to be equivalent to the Pashnev theory. This may be demonstrated similarly to the analysis in Section  \ref{Pashnev}.

There are other approaches to the Lagrangian description of massive higher-spin fields (see also \cite{Fotopoulos:2008ka} for a review and \cite{Skvortsov:2023jbn} for a recent discussion), including the powerful BRST setting of \cite{Hussain:1988uk, Pashnev:1997rm, Burdik:2000kj, Bekaert:2003uc, Francia:2010qp, Metsaev:2012uy} which has resulted in the construction of interesting models \cite{Buchbinder:2005ua, Buchbinder:2006nu, Buchbinder:2006ge, Buchbinder:2007vq}.\footnote{A unique feature of three spacetime dimensions is the existence 
of topologically massive models for higher-spin gauge fields  in both Minkowski and AdS space \cite{KP18}, which
are higher-spin extensions of linearized topologically massive gravity \cite{DJT1,DJT2}.
 They are obtained by coupling the $d=3$ counterparts of 
the massless Fronsdal  and  
Fang-Fronsdal  actions to the linearized conformal higher-spin actions. The latter are Chern-Simons-type actions
involving the linearized higher-spin Cotton tensors. In ${\mathbb M}^3$, the linearized higher-spin
Cotton tensors were constructed in the bosonic
\cite{PopeTownsend} and fermionic \cite{K16}  cases (see \cite{HHL,HLLMP} for an alternative derivation).
 In AdS$_3$, such tensors were constructed in \cite{Kuzenko:2021hyd}.
 Supersymmetric extensions of the higher-spin Cotton tensors were derived in 
\cite{K16,KT} for $\cN=1$, in \cite{KO} for $\cN=2$ and in \cite{BHHK} for  $\cN>2$ Poincar\'e supersymmetry. 
 The $\cN=1$ AdS supersymmetric extensions of the higher-spin Cotton tensors were derived in \cite{Kuzenko:2021hyd}.}
The main motivation for developing such formulations is the desire to obtain a suitable framework to address the problem of constructing consistent interactions for massive higher-spin fields within a relativistic field-theoretic framework. This problem is still open, although there have been numerous publications devoted to the construction of cubic and quartic interaction vertices, see \cite{Bekaert:2022poo} for a comprehensive list of references.
Consistent interactions of massive higher-spin fields are provided by String Theory.

It is of interest to study the relationships between different models for massive higher-spin fields. While these models lead to the 
same equations \eqref{FPequations} on the mass shell, it is important to understand whether some of these models are equivalent off the mass shell. 
The present paper is aimed at addressing (at least in part) this issue.

This paper is organized as follows. In Section \ref{Zinoviev Action} we review 
  the KZ theory in ${\mathbb M}^d$. Section \ref{Singh Hagen Correspondence} derives a $d$-dimensional extension of the Singh-Hagen theory.
  Section \ref{Pashnev} is devoted to a detailed review of the Pashnev theory \cite{Pashnev:1989}
  and its equivalence to the KZ model. Our main findings are briefly summarised in Section \ref{Section5}. 
  The main body of the paper is accompanied by two technical appendices. 
 Appendix \ref{AppendixA}, which is based on unpublished work \cite{Fegebank:2023}, is devoted to the Faddeev-Popov quantisation of the KZ model.  Appendix \ref{AppendixB}
includes additional observations in the context of the $d$-dimensional Singh-Hagen model. Specifically, we discuss alternative gauge conditions that look similar to \eqref{gauge_condition} but, unlike \eqref{gauge_condition}, do not eliminate completely the gauge freedom of the KZ model.

Throughout this paper, 
the mostly plus Minkowski metric is used, $\eta_{\m\n}=\text{diag}(-1,+1,\dots,+1)$. We denote the trace of a field $\phi^{(k)}$ with $k\geq 2$ by
\be
    \widetilde{\phi}^{(k)}_{\m_3 \dots \m_k} := \eta^{\r\s} \f^{(k)}_{\r\s \m_3\dots \m_k}~.
\ee


\section{The Klishevich-Zinoviev theory} \label{Zinoviev Action}
In this section we briefly review the KZ theory \cite{Klishevich:1997pd} for a massive spin-$s$ field in ${\mathbb M}^d$.
It  is described by the action 
\begin{subequations}  \label{Full Action for spin s}
\begin{equation}
    S_{\rm{KZ}} = \int \rd^dx\,\mathcal{L}^{(s)}~,\qquad 
      \mathcal{L}^{(s)} = \sum^{s}_{k=0}\mathcal{L}_c(\phi^{(k)})~,
 \end{equation}
where the dynamical variables $\f^{(k)} $ are symmetric double-traceless fields, and 
$  \mathcal{L}_c(\phi^{(k)}) $ has the form
\begin{equation}
    \mathcal{L}_c(\phi^{(k)}) = \mathcal{L}_0(\phi^{(k)})+\mathcal{L}_m(\phi^{(k)})~.\label{Combined Lagrangian Contribution}
\end{equation}
\end{subequations}
Here the first term on the right is Fronsdal's Lagrangian \cite{Fronsdal:1978} for a massless spin-$k$ field,
\begin{equation}
    \begin{split}
        \mathcal{L}_0(\phi^{(k)})  = & -\frac{1}{2}\partial^\mu\phi^{(k)\mu_1\dots\mu_k}\partial_\mu\phi^{(k)}_{\mu_1\dots\mu_k}+\frac{k}{2}\partial_\mu\phi^{(k)\mu\mu_2\dots\mu_k}\partial^\nu\phi^{(k)}_{\nu\mu_2\dots\mu_k}\\
        & +\frac{k(k-1)}{4}\partial^\mu\widetilde{\phi}^{(k)\mu_3\dots\mu_k}\partial_\mu\widetilde{\phi}^{(k)}_{\mu_3\dots\mu_k}+\frac{k(k-1)}{2}\partial_\mu\partial_\nu\phi^{(k)\mu\nu\mu_3\dots\mu_k}\widetilde{\phi}^{(k)}_{\mu_3\dots\mu_k}\\
        & +\frac{k(k-1)(k-2)}{8}\partial_\mu\widetilde{\phi}^{(k)\mu\mu_4\dots\mu_k}\partial^\nu\widetilde{\phi}^{(k)}_{\nu\mu_4\dots\mu_k}~.
    \end{split}\label{Massless Lagrangian Contribution}
\end{equation}
The second term in \eqref{Combined Lagrangian Contribution} is 
a massive contribution of the following structure:
\begin{equation}
    \begin{split}
        \mathcal{L}_m(\phi^{(k)})  = & a_k\phi^{(k-1)\mu_2\dots\mu_{k}}\partial^\mu\phi^{(k)}_{\mu\mu_2\dots\mu_k}+b_k\widetilde{\phi}^{(k)\mu_3\dots\mu_k}\partial^\mu\phi^{(k-1)}_{\mu\mu_3\dots\mu_k}\\
        & +c_k\partial_\mu\widetilde{\phi}^{(k)\mu\mu_4\dots\mu_k}\widetilde{\phi}^{(k-1)}_{\mu_4\dots\mu_k}+d_k\phi^{(k)\mu_1\dots\mu_k}\phi^{(k)}_{\mu_1\dots\mu_k}\\
        & +e_k\widetilde{\phi}^{(k)\mu_3\dots\mu_k}\widetilde{\phi}^{(k)}_{\mu_3\dots\mu_k}-f_k\widetilde{\phi}^{(k)\mu_3\dots\mu_k}\phi^{(k-2)}_{\mu_3\dots\mu_k}~.
    \end{split}\label{k^th Massive Lagrangian Contribution}
\end{equation}
It involves several numerical coefficients, all of which are determined by requiring the action \eqref{Full Action for spin s} to be invariant under gauge transformations 
\begin{equation}
    \begin{split}
        \delta\phi^{(k)}_{\mu_1\dots\mu_k} =~ & \partial_{(\mu_1}\xi^{(k-1)}_{\mu_2\dots\mu_k)}+\alpha_k\xi^{(k)}_{\mu_1\dots\mu_k}-\frac{k(k-1)}{2}\beta_k\eta_{(\mu_1\mu_2}\xi^{(k-2)}_{\mu_3\dots\mu_k)} ~,
    \end{split}\label{k^th Gauge Transformation}
\end{equation}
where $\a_k$ and $\b_k$ are some coefficients,
and $\xi^{(k)}$ are symmetric and traceless gauge parameters.\footnote{As usual, it is assumed 
in \eqref{k^th Gauge Transformation} that when a term would contain fields of negative rank in the above, that term is ignored. }
Direct calculations show that  the coefficients in \eqref{k^th Massive Lagrangian Contribution} 
and $\b$'s in \eqref{k^th Gauge Transformation} are fixed in terms of $\a$'s:
\begin{subequations}\label{Zinoviev Coefficients}
\bea
        a_k &=& -k\alpha_{k-1}~,\hspace{20pt} b_k = -k(k-1)\alpha_{k-1}~, \hspace{20pt}c_k = -\frac{k(k-1)(k-2)}{4}\alpha_{k-1}~,\\
             2d_k &=& \frac{2(k+1)(2k+d-3)}{2k+d-4}(\alpha_{k})^2-k(\alpha_{k-1})^2~, 
        \quad        k\geq1~; \qquad  d_0 = \frac{d}{d-2}(\alpha_1)^2~, \\
        2e_k &=& -\frac{k(k^2-1)(2k+d)}{4(2k+d-4)}(\alpha_k)^2+\frac{k^2(k-1)}{2}(\alpha_{k-1})^2~,\\
        2f_k &=& -k(k-1)\alpha_{k-1}\alpha_{k-2}~, \\
        \beta_k &=& \frac{2\alpha_{k-1}}{(k-1)(2k+d-6)}~.
\eea
\end{subequations}
For the coefficients $\a$'s one gets
\begin{subequations} \label{2.6}
\bea
(\alpha_k)^2 = \frac{s(s-k)(s+k+d-3)}{(k+1)(2k+d-2)}(\a_{s-1})^2~, \hspace{20pt} 0\leq k\leq s-2~,
\eea
for which it is convenient to choose $(\a_{s-1})^2$:
\bea
\label{Alpha k}
    (\alpha_{s-1})^2 = \frac{m^2}{s}\quad \implies \quad
    (\alpha_k)^2 = \frac{(s-k)(s+k+d-3)}{(k+1)(2k+d-2)}m^2~, \hspace{20pt} 0\leq k\leq s-2~.
\eea
\end{subequations}


\section{Singh-Hagen model in $d$ dimensions}
\label{Singh Hagen Correspondence}

In four dimensions, it was shown by Klishevich and Zinoviev \cite{Klishevich:1997pd} that the Singh-Hagen theory 
 \cite{Singh:1974}  is obtained from the $d=4$ version of the theory described in the previous section by appropriately fixing the gauge freedom.
 Here we extend the analysis of \cite{Klishevich:1997pd} to $d$ dimensions. 

In order to derive a $d$-dimensional analogue of the Singh-Hagen theory from the KZ one, it suffices to choose a unitary gauge. To start with, for $k>1$ we decompose each double-traceless symmetric field $\f^{(k)}$ into a sum of two traceless symmetric fields
$\omega^{(k)}$ and 
${\varphi}^{(k-2)}$. In terms of these, the field  $\f^{(k)}$ is
\begin{equation}
    \phi^{(k)}_{\mu_1\dots\mu_k}=\omega^{(k)}_{\mu_1\dots\mu_k}+
       \frac{k(k-1)}{2(d+2k-4)}
    \eta_{(\mu_1\mu_2}{\varphi}^{(k-2)}_{\mu_3\dots\mu_k)}~,
\end{equation}
with  $\widetilde{\f}^{(k)}_{\mu_1\dots\mu_{k-2}} = {\varphi}^{(k-2)}_{\mu_1\dots\mu_{k-2}}$, with $k=s, \dots ,2$.
Then, the gauge freedom \eqref{k^th Gauge Transformation} may be completely fixed by imposing the conditions 
\bea
\omega^{(k)}_{\mu_1\dots\mu_k} =0~, \qquad k = 0, 1, \dots , s-1~.
\label{gauge_condition}
\eea
The remaining rank-$s$ field will be re-labelled as follows: $\omega^{(s)}_{\mu_1\dots\mu_s} =\varphi^{(s)}_{\mu_1\dots\mu_s}$.
As a result, upon imposing the gauge condition \eqref{gauge_condition} we stay with the following fields
\begin{subequations} \label{unitary}
\bea
\phi^{(s)}_{\mu_1\dots\mu_s}&=&\varphi^{(s)}_{\mu_1\dots\mu_s}+\frac{s(s-1)}{2(d+2s-4)}\eta_{(\mu_1\mu_2}\varphi^{(s-2)}_{\mu_3\dots\mu_s)}~,\\
  \phi^{(k)}_{\mu_1\dots\mu_k}&=&
     \frac{k(k-1)}{2(d+2k-4)}
  \eta_{(\mu_1\mu_2}\varphi^{(k-2)}_{\mu_3\dots\mu_k)}~, \qquad 
  k\leq s-1~.
\eea
\end{subequations}

We now turn to massaging the separate contributions to the Lagrangian 
$\mathcal{L}^{(s)}$, eq. 
\eqref{Full Action for spin s}, in the gauge \eqref{gauge_condition} 
or, equivalently, \eqref{unitary}.
For the massless Lagrangian $ \mathcal{L}_0(\phi^{(s)}) $ we obtain
\begin{align}
\label{Massless Spin s Lagrangian Varphi}
        \mathcal{L}_0(\phi^{(s)}) 
             = & \frac{1}{2}\varphi^{(s)\mu_1\dots\mu_s}\square\varphi^{(s)}_{\mu_1\dots\mu_s}
        +\frac{s}{2}\partial_\mu\varphi^{(s)\mu\mu_2\dots\mu_s}\partial^\nu\varphi^{(s)}_{\nu\mu_2\dots\mu_s} \non \\
&        +\frac{s(s-1)(d+2s-6)}{2(d+2s-4)}\partial_\mu\partial_\nu\varphi^{(s)\mu\nu\mu_3\dots\mu_s}\varphi^{(s-2)}_{\mu_3\dots\mu_s} \non \\
        & -\frac{s(s-1)(d+2s-6)(d+2s-5)}{4(d+2s-4)^2}\varphi^{(s-2)\mu_3\dots\mu_s}\square\varphi^{(s-2)}_{\mu_3\dots\mu_s}\\
        & +\frac{(s-2)(d+2s-6)(d+2s-8)s(s-1)}{8(d+2s-4)^2}\partial_\mu\varphi^{(s-2)\mu\mu_4\dots\mu_s}\partial^\nu\varphi^{(s-2)}_{\nu\mu_4\dots\mu_s}~.\non
\end{align}
For $0\leq k\leq s-1$, the massless Lagrangian $ \mathcal{L}_0(\phi^{(k)})  $ leads to 
\begin{align}
        \mathcal{L}_0(\phi^{(k)})  = & -\frac{k(k-1)(d+2k-6)(d+2k-5)}{4(d+2k-4)^2}\varphi^{(k-2)\mu_3\dots\mu_k}\square\varphi^{(k-2)}_{\mu_3\dots\mu_k} \non \\
        &+\frac{k(k-1)(k-2)(d+2k-6)(d+2k-8)}{8(d+2k-4)^2}\partial_\mu\varphi^{(k-2)\mu\mu_4\dots\mu_k}\partial^\nu\varphi^{(k-2)}_{\nu\mu_4\dots\mu_k}~.
\end{align}
Next, we turn to the massive contributions 
\eqref{k^th Massive Lagrangian Contribution}.
For $0\leq k \leq s-1$, we obtain
\begin{align}
        \mathcal{L}_m(\phi^{(k)})  = & a_k\frac{(k-1)(k-2)}{2(d+2k-6)}\varphi^{(k-3)\mu_4\dots\mu_k}\partial^\mu\varphi^{(k-2)}_{\mu\mu_4\dots\mu_k}+b_k\frac{k-2}{d+2k-6}\varphi^{(k-2)\mu\mu_4\dots\mu_k}\partial_\mu\varphi^{(k-3)}_{\mu_4\dots\mu_k} \non \\
        &+c_k\partial_\mu\varphi^{(k-2)\mu\mu_4\dots\mu_k}\varphi^{(k-3)}_{\mu_4\dots\mu_k}+d_k\frac{k(k-1)}{2(d+2k-4)}\varphi^{(k-2)\mu_3..\mu_k}\varphi^{(k-2)}_{\mu_3..\mu_k}\\
        & +e_k\varphi^{(k-2)\mu_3\dots\mu_k}\varphi^{(k-2)}_{\mu_3\dots\mu_k}~.
        \non
\end{align}
Substituting in the values for $a_k$, $b_k$ and $c_k$ from (\ref{Zinoviev Coefficients}), 
upon integration by parts we get
\begin{align}
        \mathcal{L}_m(\phi^{(k)})  = & \alpha_{k-1}\frac{k(k-1)(k-2)(d+2k-8)}{4(d+2k-6)}\varphi^{(k-3)\mu_4\dots\mu_k}\partial^\mu\varphi^{(k-2)}_{\mu\mu_4\dots\mu_k}\non \\
        & +\left(d_k\frac{k(k-1)}{2(d+2k-4)}+e_k\right)\varphi^{(k-2)\mu_3..\mu_k}\varphi^{(k-2)}_{\mu_3..\mu_k}~.
 \end{align}
On the other hand, the massive contribution $ \mathcal{L}_m(\phi^{(s)})  $ is
\begin{align}
\label{Massive Spin s Lagrangian Varphi}
           \mathcal{L}_m(\phi^{(s)})  = & d_s\varphi^{(s)\mu_1\dots\mu_s}\varphi^{(s)}_{\mu_1..\mu_s}+\alpha_{s-1}\frac{s(s-1)(s-2)(d+2s-8)}{4(d+2s-6)}\varphi^{(s-3)\mu_4\dots\mu_s}\partial^\mu\varphi^{(s-2)}_{\mu\mu_4\dots\mu_s} \non \\
        & +\left(d_s\frac{s(s-1)}{2(d+2s-4)}+e_s\right)\varphi^{(s-2)\mu_3..\mu_s}\varphi^{(s-2)}_{\mu_3..\mu_s}~.
   \end{align}

Finally, it remains to substitute the contributions
 (\ref{Massless Spin s Lagrangian Varphi}--\ref{Massive Spin s Lagrangian Varphi}) 
into the  full Lagrangian (\ref{Full Action for spin s}) as well as to make use of \eqref{Zinoviev Coefficients}
and \eqref{2.6}
to end up with 
\begin{align}
    \mathcal{L}^{(s)}_{\rm SH} = &\frac{1}{2}\varphi^{(s)\mu_1\dots\mu_s}(\square-m^2)\varphi^{(s)}_{\mu_{1}\dots\mu_{s}}+\frac{1}{2}s\partial^\nu\varphi^{(s)}_{\nu\mu_2\dots\mu_s}\partial_\lambda\varphi^{(s)\lambda\mu_2\dots\mu_s}\non \\
    &+\frac{s(s-1)(d+2s-6)}{2(d+2s-4)}\partial_\mu\partial_\nu\varphi^{(s)\mu\nu\mu_3\dots\mu_{s}}\varphi^{(s-2)}_{\mu_3\dots\mu_{s}}\non \\
    &-\frac{s(s-1)(d+2s-6)(d+2s-5)}{4(d+2s-4)^2}\varphi^{(s-2)\mu_3\dots\mu_{s}}\left(\square-\frac{d+2s-4}{d+2s-6}m^2\right)\varphi^{(s-2)}_{\mu_3\dots\mu_{s}} \non \\
    & +\frac{s(s-1)(s-2)(d+2s-6)(d+2s-8)}{8(d+2s-4)^2}\partial_\mu\varphi^{(s-2)\mu\mu_4\dots\mu_{s}}\partial^\nu\varphi^{(s-2)}_{\nu\mu_4\dots\mu_{s}}\non \\
    &-\sum^s_{k=3}\left(\frac{(s-k+2)(s-k+1)(d+2s-2k-2)}{4(d+2s-2k)}\right)
    \label{d-dimSH}
    \\
    & \times\left[\frac{(d+2s-2k-1)}{(d+2s-2k)}\varphi^{(s-k)\mu_{k+1}\dots\mu_s}\left(\square-\frac{k(d+2s-2k)(d+2s-k-3)}{2(d+2s-2k-2)(d+2s-2k-1)}m^2\right)\varphi^{(s-k)}_{\mu_{k+1}\dots\mu_s}\right.\non \\
    & \hspace{20pt}-\frac{(s-k)(d+2s-2k-4)}{2(d+2s-2k)}\partial_\mu\varphi^{(s-k)\mu\mu_{k+2}\dots\mu_s}\partial^\nu\varphi^{(s-k)}_{\nu\mu_{k+2}\dots\mu_s}\non \\
    & \left.\hspace{20pt}-\sqrt{\frac{(k-2)(s-k+3)(d+2s-k-1)}{(d+2s-2k+2)}}m\varphi^{(s-k)\mu_{k+1}\dots\mu_s}\partial^\mu\varphi^{(s-k+1)}_{\mu\mu_{k+1}\dots\mu_s}\right]~.\non 
\end{align}
This Lagrangian defines the Singh-Hagen model in $d$ dimensions, which has so far been described in the literature only in the $d=4$ case \cite {Singh:1974}, see below. This $d$-dimensional extension of the Singh-Hagen model turns out to be unique in the sense given in Appendix \ref{AppendixB}.

Let us analyse the equation of motion corresponding to \eqref{d-dimSH}.
It is useful to adopt the  Singh-Hagen notation 
$\{ \dots\}_{\rm S.T.}$, which denotes the symmetric and traceless component of the term within the brackets.
The equation of motion for the field $\varphi^{(s)}$ is
\bsubeq
\begin{align}
    &(-\square+m^2)\varphi^{(s)}_{\mu_1\dots\mu_s}+s\partial^\nu\{\partial_{\mu_1}\varphi^{(s)}_{\nu\mu_2\dots \mu_s}\}_{\rm S.T.}=\frac{s(s-1)(d+2s-6)}{2(d+2s-4)}\{\partial_{\mu_1}\partial_{\mu_2}\varphi^{(s-2)}_{\mu_3\dots\mu_s}\}_{\rm S.T.}\label{SH d-dim EoM (s)}~.
 \end{align} 
The equation of motion for the field $\varphi^{(s-2)}$ is    
\begin{align}
  &\frac{s(s-1)(d+2s-6)(d+2s-5)}{(d+2s-4)^2}\left(\square-\frac{d+2s-4}{d+2s-6}m^2\right)\varphi^{(s-2)}_{\mu_3\dots\mu_{s}}\non \\
    &+\frac{s(s-1)(s-2)(d+2s-6)(d+2s-8)}{2(d+2s-4)^2}\partial^\nu\{\partial_{\mu_3}\varphi^{(s-2)}_{\nu\mu_4\dots\mu_s}\}_{\rm S.T.} \label{SH d-dim EoM (s-2)}\\
    &+\frac{(s-1)(s-2)(d+2s-4)\sqrt{s}}{2(d+2s-6)}m\{\partial_{\mu_3}\varphi^{(s-3)}_{\mu_4\dots\mu_s}\}_{\rm S.T.}
    =\frac{s(s-1)(d+2s-6)}{(d+2s-4)}\partial^\nu\partial^\lambda\varphi^{(s)}_{\nu\lambda\mu_3\dots\mu_s}~. \non
\end{align}
Finally, the equations of motion for the fields $\varphi^{(s-k)}$, with $3\leq k \leq s$, are the following:
\begin{align}
    & \qquad \left(\frac{(s-k+2)(s-k+1)(d+2s-2k-2)}{4(d+2s-2k)}\right)\non \\
    & \times\left[\frac{2(d+2s-2k-1)}{(d+2s-2k)}\left(\square-\frac{k(d+2s-2k)(d+2s-k-3)}{2(d+2s-2k-2)(d+2s-2k-1)}m^2\right)\varphi^{(s-k)}_{\mu_{k+1}\dots \mu_s}\right.\non \\
    & \hspace{20pt}+\frac{(s-k)(d+2s-2k-4)}{(d+2s-2k)}\partial^\nu\{\partial_{\mu_{k+1}}\varphi^{(s-k)}_{\nu\mu_{k+2}\dots\mu_s}\}_{\rm S.T.}\label{SH d-dim EoM (s-q)} \\
    & \left.\hspace{20pt}-\sqrt{\frac{(k-2)(s-k+3)(d+2s-k-1)}{(d+2s-2k+2)}}m\partial^\mu\varphi^{(s-k+1)}_{\mu\mu_{k+1}\dots\mu_s}\right] \non\\
    =&  -\left(\frac{(s-k+3)(s-k+2)(d+2s-2k)}{4(d+2s-2k+2)}     \right) \non \\
&\times     \sqrt{\frac{(k-3)(s-k+4)(d+2s-k)}{(d+2s-2k)}}m\{\partial_{\mu_{k+1}}\varphi^{(s-k-1)}_{\mu_{k+2}\dots\mu_s}\}_{\rm S.T.}~.\non
\end{align}
\esubeq
After some algebra, it may be seen that these equations yield $\varphi^{(s-k)}=0$ for $s\geq k\geq 2$, while 
the field $\varphi^{(s)}$ obeys the Fierz-Pauli equations
 \eqref{Fierz Pauli}. As a result, the number of on-shell degrees of freedom is 
 \bea
   n(d,s)  =   \frac{2s+d-3}{d-3}\binom{d+s-4}{s}~,
   \label{3.11}
\eea
which reduces to $2s+1$ degrees of freedom in four spacetime dimensions.  In three spacetime dimensions, 
\eqref{3.11} should be replaced with $n(3,s) =2$. 

As an instructive check, it is worth 
comparing \eqref{d-dimSH} directly with the Singh-Hagen model
 \cite {Singh:1974} formulated in ${\mathbb M}^4$.
Choosing $d=4$ in \eqref{d-dimSH} yields
\begin{align}
    \mathcal{L} = &\frac{1}{2}\varphi^{(s)\mu_1\dots\mu_s}(\square-m^2)\varphi^{(s)}_{\mu_{1}\dots\mu_{s}}+\frac{1}{2}s\partial^\nu\varphi^{(s)}_{\nu\mu_2\dots\mu_s}\partial_\lambda\varphi^{(s)\lambda\mu_2\dots\mu_s}+\frac{(s-1)^2}{2}\partial_\mu\partial_\nu\varphi^{(s)\mu\nu\mu_3\dots\mu_{s}}\varphi^{(s-2)}_{\mu_3\dots\mu_{s}}\non \\
    &-\frac{(s-1)^2(2s-1)}{8s}\varphi^{(s-2)\mu_3\dots\mu_{s}}\left(\square-\frac{s}{s-1}m^2\right)\varphi^{(s-2)}_{\mu_3\dots\mu_{s}}\non \\
    & +\frac{(s-1)^2(s-2)^2}{8s}\partial_\mu\varphi^{(s-2)\mu\mu_4\dots\mu_{s}}\partial^\nu\varphi^{(s-2)}_{\nu\mu_4\dots\mu_{s}}\\
    &-\sum^s_{k=3}\left[\frac{(s-k+1)^2(2s-2k+3)}{8(s-k+2)}\varphi^{(s-k)\mu_{k+1}\dots\mu_s}\left(\square-\frac{k(s-k+2)(2s-k+1)}{2(s-k+1)(2s-2k+3)}m^2\right)\varphi^{(s-k)}_{\mu_{k+1}\dots\mu_s}\right.\non \\
    & \hspace{30pt}-\frac{(s-k+1)^2(s-k)^2}{8(s-k+2)}\partial_\mu\varphi^{(s-k)\mu\mu_{k+2}\dots\mu_s}\partial^\nu\varphi^{(s-k)}_{\nu\mu_{k+2}\dots\mu_s}\non \\
    & \left.\hspace{30pt}-\frac{(s-k+1)^2}{4}\sqrt{\frac{(k-2)(2s-k+3)}{2}}m\varphi^{(s-k)\mu_{k+1}\dots\mu_s}\partial^\mu\varphi^{(s-k+1)}_{\mu\mu_{k+1}\dots\mu_s}\right]~.\non 
\end{align}
Let us rescale the fields $\varphi^{(s-2)}$ and $\varphi^{(s-k)}$, with $3\leq k\leq s$, as follows:
\bsubeq
\bea
    \varphi^{(s-2)}_{\mu_{3}\dots\mu_s}&\longrightarrow &\frac{2s}{2s-1}\varphi^{(s-2)}_{\mu_{3}\dots\mu_s}~, \\
    \varphi^{(s-k)}_{\mu_{k+1}\dots\mu_s}&\longrightarrow& \frac{2(s-1)}{(s-k+1)}\sqrt{\frac{s(s-k+2)}{(2s-1)(2s-2k+3)}}
    \non\\
    &&\times\left(\prod^{k-1}_{n=2}\sqrt{\frac{(n-1)(s-n)^2(s-n+2)(2s-n+2)}{2(s-n+1)(2s-2n+1)(2s-2n+3)}}\right)\varphi^{(s-k)}_{\mu_{k+1}\dots\mu_s}~.
\eea
\esubeq
This yields
\begin{align}
    \mathcal{L} = &\frac{1}{2}\varphi^{(s)\mu_1\dots\mu_s}(\square-m^2)\varphi^{(s)}_{\mu_{1}\dots\mu_{s}}+\frac{1}{2}s\partial^\nu\varphi^{(s)}_{\nu\mu_2\dots\mu_s}\partial_\lambda\varphi^{(s)\lambda\mu_2\dots\mu_s}\non \\
    &+\frac{s(s-1)^2}{2s-1}\left\{\partial_\mu\partial_\nu\varphi^{(s)\mu\nu\mu_3\dots\mu_{s}}\varphi^{(s-2)}_{\mu_3\dots\mu_{s}}-\frac{1}{2}\varphi^{(s-2)\mu_3\dots\mu_{s}}\left(\square-\frac{s}{s-1}m^2\right)\varphi^{(s-2)}_{\mu_3\dots\mu_{s}}\right.\non \\
    & +\frac{(s-2)^2}{2(2s-1)}\partial_\mu\varphi^{(s-2)\mu\mu_4\dots\mu_{s}}\partial^\nu\varphi^{(s-2)}_{\nu\mu_4\dots\mu_{s}}\\
    &-\sum^s_{k=3}\left(\prod^{k-1}_{n=2}\frac{(n-1)(s-n)^2(s-n+2)(2s-n+2)}{2(s-n+1)(2s-2n+1)(2s-2n+3)}\right)\non \\
    & \times\left[\frac{1}{2}\varphi^{(s-k)\mu_{k+1}\dots\mu_s}\left(\square-\frac{k(s-k+2)(2s-k+1)}{2(s-k+1)(2s-2k+3)}m^2\right)\varphi^{(s-k)}_{\mu_{k+1}\dots\mu_s}\right.\non \\
    & \hspace{15pt} \left.\left.-\frac{(s-k)^2}{2(2s-2k+3)}\partial_\mu\varphi^{(s-k)\mu\mu_{k+2}\dots\mu_s}\partial^\nu\varphi^{(s-k)}_{\nu\mu_{k+2}\dots\mu_s}-m\varphi^{(s-k)\mu_{k+1}\dots\mu_s}\partial^\mu\varphi^{(s-k+1)}_{\mu\mu_{k+1}\dots\mu_s}\right]\right\}~,\non 
\end{align}
which is exactly the Lagrangian derived by  Singh and Hagen  \cite {Singh:1974}.

Extensions of the Singh-Hagen model to $d$ dimensions have previously been considered in \cite{Lindwasser,Buchbinder:2015, Buchbinder:2022lsi}. We provide comment on these extensions in subsection \ref{Pashnev Gauge Fixing}.


\section{Variations on the Pashnev theory}\label{Pashnev}
The KZ theory is not the first gauge-invariant formulation for a massive higher-spin field.  
Perhaps the earliest such model was derived by Pashnev \cite{Pashnev:1989} by dimensional reduction of (\ref{Massless Lagrangian Contribution}).
The latter theory is formulated in terms of of four independent symmetric and traceful fields of rank $s$, $s-1$, $s-2$ and $s-3$, while the gauge parameters
are described by  two independent symmetric and traceful fields with rank $s-1$ and $s-2$.  For this reason it is often referred to as the quartet formulation.

In this section we study a correspondence between the KZ and Pashnev theories. We start with an extended  review of \cite{Pashnev:1989}.


\subsection{Oscillator formalism} \label{Oscillator Formalism}
Following \cite{Pashnev:1989}, we first introduce the so-called oscillator formalism. 
Consider a Fock space with bosonic oscillators $a_\mu$ and $a_\mu^+$ whose commutator is
\bea
\label{Ladder Commutator}
    [a_\mu,a_\nu^+]=\eta_{\mu\nu}~.
\eea
Associated with a complex rank-$s$ symmetric field $\f^{(s)}$ is the ket-state
\be
    |\f^{(s)}\rangle:=\frac{1}{s!}\f^{(s)\mu_1\dots\m_s}a^+_{\m_1}\dots a^+_{\m_s}|0\rangle~,
\ee
while the bra-state
\be
    \langle\f^{(s)}|:=\frac{1}{s!}\langle0|a_{\m_1}\dots a_{\m_s}\overline{\f}^{(s)\mu_1\dots\m_s}
\ee
is associated with the conjugate $\overline{\f}^{(s)}$ of  $\f^{(s)}$.
Here $|0\rangle$ denotes the the Fock  vacuum state,
\be
    a^\mu|0\rangle =0~.
\ee

We introduce the Hermitian operator
\bsubeq
\label{Operators}
\bea
    l_0:= p^\m p_\m~,  \qquad p_\m=-\ri\partial_\m~,
\eea
as well as non-Hermitian operators
\be
    l_1:= p^\m a_\m~, \qquad l_2:=\frac{1}{2}a^\m a_\m~,
\ee
and their conjugates
\be
    l_1^+:= p^\m a^+_\m~, \qquad l^+_2:=\frac{1}{2}a^{+\m} a^+_\m~.
\ee
\esubeq
The algebra of these operators is given by
\bea
\label{Algebra Fronsdal}
    \left[l_{1},l_{1}^+\right]=l_0~,\qquad \left[l_{1},l_{2}^+\right]=l^+_{1}~,\qquad \left[l^+_{1},l_{2}\right]=-l_{1}~, \qquad \left[l_{2},l_{2}^+\right]=g_0~, \non\\
    \left[l_{1},g_0\right]=l_1~,\qquad \left[l_1^+,g_0\right]=-l_1^+~,\qquad \left[l_{2},g_0\right]=2l_{2}~,\qquad \left[l_{2}^+,g_0\right]=-2l_{2}^+~.
\eea
Here we have introduced the operator 
\be
    g_0 = a^{+\m}a_\m+\frac{d}{2}~,
\ee
which encodes the spin of a state $|\phi^{(s)}\rangle $ via its eigenvalue
\be
    g_0|\phi^{(s)}\rangle = \left(s+\frac{d}{2}\right)|\f^{(s)}\rangle~.
\ee
The actions of \eqref{Operators} on a ket-state correspond to actions on the tensor field as follows
\bsubeq
\be
    l_0|\f^{(s)}\rangle \rightarrow -\square\f^{(s)}_{\m\dots\m_s}~, \qquad l_1|\f^{(s)}\rangle\rightarrow -\ri\partial^\m\f^{(s)}_{\m\m_2\dots\m_s}~, \qquad l_2|\f^{(s)}\rangle\rightarrow \frac{1}{2}\widetilde{\f}^{(s)}_{\m_3\dots\m_s}~,
\ee
while
\be
    l^+_1|\f^{(s)}\rangle\rightarrow -\ri\partial_{(\m}\f^{(s)}_{\m_1\dots\m_s)}~, \qquad l_2^+|\f^{(s)}\rangle\rightarrow \frac{1}{2}\eta_{(\m\n}\f^{(s)}_{\m_1\dots\m_s)}~.
\ee
On a bra-state, their actions instead correspond to
\be
    \langle\f^{(s)}|l_0 \rightarrow -\square\overline{\f}^{(s)}_{\m\dots\m_s}~, \qquad \langle\f^{(s)}|l_1\rightarrow \ri\partial_{(\m}\overline{\f}^{(s)}_{\m_1\dots\m_s)}~, \qquad \langle\f^{(s)}|l_2\rightarrow \frac{1}{2}\eta_{(\m\n}\overline{\f}^{(s)}_{\m_1\dots\m_s)}~,
\ee
while
\be
    \langle\f^{(s)}|l^+_1\rightarrow \ri\partial^\m\overline{\f}^{(s)}_{\m\m_2\dots\m_s}~, \qquad \langle\f^{(s)}|l_{2}^+\rightarrow \frac{1}{2}\widetilde{\overline{\f}}^{(s)}_{\m_3\dots\m_s}~.
\ee
\esubeq

Let us consider a field which transforms via
\be
\label{Gauge Transformation Oscillator Massless}
    \d|\f^{(s)}\rangle = \ri l^+_1|\x^{(s-1)}\rangle~,
\ee
where $|\x^{(s-1)}\rangle$ is the ket-state associated with a rank-$(s-1)$ symmetric gauge parameter $\x^{(s-1)}$. We impose a double-traceless condition on the $\f^{(s)}$ field and a traceless condition on the $\x^{(s-1)}$ gauge parameter. In Fock space, these are expressed by the conditions
\bsubeq
\label{Oscillator Tracelessness}
\be
    (l_{2})^2|\f^{(s)}\rangle = 0~.
\ee
and
\be
    l_{2}|\x^{(s-1)}\rangle = 0~,
\ee
\esubeq
respectively.
The free-field Lagrangian whose action is invariant under (\ref{Gauge Transformation Oscillator Massless}) is
\bea
\label{Fronsdal Action Oscillator Formalism}
    \mathcal{L}\,(\f^{(s)}) &=& \frac{1}{2}\langle\f^{(s)}|\left(-l_0+l_1^+l_1+2l_2^+l_0 l_2-(l_1^+)^2l_2-l_2^+(l_1)^2+l_2^+l_1^+l_1l_2\right)|\f^{(s)}\rangle~.
\eea
We easily see that under the reality conditions
\be
    \f^{(s)}_{\m_1\dots\m_s} = \overline{\f}^{(s)}_{\m_1\dots\m_s}
\ee
and
\be
    \x^{(s-1)}_{\m_1\dots\m_s} = \overline{\x}^{(s-1)}_{\m_1\dots\m_{s-1}}~,
\ee
we get an exact correspondence with the Fronsdal action for real massless fields. Then (\ref{Fronsdal Action Oscillator Formalism}) describes complex massless fields of spin $s$ in oscillator formalism. By considering a dimensional reduction on this complex field theory rather than the real field theory, we can make a different choice of reality conditions which will prove more convenient for the massive theory.

\subsection{Dimensional reduction of Fronsdal's theory} \label{Dim Reduction} 
We consider now the dimensional reduction of the theory outlined in the previous subsection for fields and gauge parameters in $\mathbb{M}^{d+1}$. We substitute
\bsubeq
\label{Substitution}
\be
    |\phi^{(s)}\rangle=\re^{\ri x_{d}m}|\check{\phi}^{(s)}\rangle~,
\ee
and
\be
    |\xi^{(s-1)}\rangle=\re^{\ri x_{d}m}|\check{\xi}^{(s-1)}\rangle
\ee
\esubeq
where $x_{d}$ is the $(d+1)^{\text{th}}$ spacetime coordinate, while $|\check{\phi}^{(s)}\rangle$ and $|\check{\xi}^{(s-1)}\rangle$ are ket-states corresponding to the fields and gauge parameters in $\mathbb{M}^d$. The effect is that the momentum operator $p_d$ is replaced with $m$. Defining
\be
    \check{l}_{1}^{(+)}:=l^{(+)}_{1}+mb^{(+)}~,\qquad \check{l}_{2}^{(+)}:=l^{(+)}_2+\frac{(b^{(+)})^2}{2}~,
\ee
where $l_{1}$, $l_{2}$ are hereby taken to be explicitly the $d$-dimensional operators, while $b:=a_d$\footnote{It is obvious from (\ref{Ladder Commutator}) that $[b,b^+]=1$.}, and 
\be
    l_0:=-\square+m^2~,
\ee
the algebra (\ref{Algebra Fronsdal}) is
\bea
\label{Algebra Dim Reduction}
    \left[\check{l}_{1},\check{l}_{1}^+\right]=l_0~,\qquad \left[\check{l}_{1},\check{l}_{2}^+\right]=\check{l}^+_{1}~,\qquad \left[\check{l}^+_{1},\check{l}_{2}\right]=-\check{l}_{1}~, \qquad \left[\check{l}_{2},\check{l}_2^+\right]=g_0~, \non\\
    \left[\check{l}_{1},g_0\right]=\check{l}_1~,\qquad \left[\check{l}_1^+,g_0\right]=-\check{l}_1^+~,\qquad \left[\check{l}_2,g_0\right]=2\check{l}_2~,\qquad \left[\check{l}_2^+,g_0\right]=-2\check{l}_2^+~,
\eea
where the operator $g_0$ is taken as
\be
    g_0 = a^{+\m}a_\m+b^+b+\frac{d+1}{2}~.
\ee
The conditions (\ref{Oscillator Tracelessness}) become
\bsubeq
\label{Traceless Dim Reduced}
\be
\label{Phi Traceless Dim Reduced}
    (\check{l}_2)^2|\check{\f}^{(s)}\rangle = 0~,
\ee
and
\be
\label{Xi Traceless Dim Reduced}
    \check{l}_2|\check{\x}^{(s-1)}\rangle = 0~,
\ee
\esubeq
and the Lagrangian (\ref{Fronsdal Action Oscillator Formalism}) is now given by
\bea
\label{Fronsdal Action Dim Reduced}
    \mathcal{L}(\check{\f}^{(s)}) &=& \frac{1}{2}\langle\check{\f}^{(s)}|\left(-l_0+\check{l}_1^+\check{l}_1+2\check{l}_2^+l_0 \check{l}_{2}-(\check{l}_1^+)^2\check{l}_{2}-\check{l}_{2}^+(\check{l}_1)^2+\check{l}_{2}^+\check{l}_1^+\check{l}_1\check{l}_{2}\right)|\check{\f}^{(s)}\rangle~,
\eea
with the associated gauge transformation (\ref{Gauge Transformation Oscillator Massless}) becoming
\be
\label{Gauge Transformation Dim Reduced}
    \d|\check{\f}^{(s)}\rangle=\ri \check{l}_1^+|\check{\x}^{(s-1)}\rangle~.
\ee
Since $|\f^{(s)}\rangle$ describes a spin-$s$ massless field in $d+1$ dimensions, we require that 
\be
    g_0|\f^{(s)}\rangle = \left(s+\frac{d+1}{2}\right)|\f^{(s)}\rangle~, \qquad g_0|\x^{(s-1)}\rangle = \left(s-1+\frac{d+1}{2}\right)|\x^{(s-1)}\rangle~.
\ee
It then follows from (\ref{Substitution}) that
\be
    g_0|\check{\f}^{(s)}\rangle = \left(s+\frac{d+1}{2}\right)|\check{\f}^{(s)}\rangle~, \qquad g_0|\check{\x}^{(s-1)}\rangle = \left(s-1+\frac{d+1}{2}\right)|\check{\x}^{(s-1)}\rangle~,
\ee
and thus we construct
\bsubeq
\be
\label{Phi Field}
    |\check{\f}^{(s)}\rangle=\sum_{k=0}^{s}\frac{(b^+)^k}{k!}| \psi^{(s-k)}\rangle~,
\ee
as well as
\be
\label{Xi Field}
    |\check{\x}^{(s-1)}\rangle=\sum_{k=0}^{s-1}\frac{(b^+)^k}{k!}|\ve^{(s-k-1)}\rangle~,
\ee
\esubeq
where $| \psi^{(k)}\rangle$ and $|\ve^{(k)}\rangle$ correspond to traceful states in $d$ dimensions which are independent of $b$ and $b^+$. Inserting (\ref{Phi Field}) into (\ref{Phi Traceless Dim Reduced}) leads to the relations
\bsubeq
\label{traceful Fields}
\be
    | \psi^{(s-2k)}\rangle = (-1)^{k+1}\left[k(2l_{2})^{k-1}| \psi^{(s-2)}\rangle+(k-1)(2l_{2})^{k}| \psi^{(s)}\rangle\right]~,
\ee
and
\be
    | \psi^{(s-2k-1)}\rangle = (-1)^{k+1}\left[k(2l_{2})^{k-1}| \psi^{(s-3)}\rangle+(k-1)(2l_{2})^{k}| \psi^{(s-1)}\rangle\right]~,
\ee
\esubeq
and hence the Lagrangian depends only on four independent and traceful fields corresponding to the states $| \psi^{(s)}\rangle$, $| \psi^{(s-1)}\rangle$, $| \psi^{(s-2)}\rangle$ and $| \psi^{(s-3)}\rangle$. Doing the same with equations (\ref{Xi Field}) and (\ref{Xi Traceless Dim Reduced}) gives
\bsubeq
\label{traceful Gauge}
\be
    |\ve^{(s-2k-1)}\rangle = (-1)^{k}(2l_{2})^k|\ve^{(s-1)}\rangle~,
\ee
and
\be
    |\ve^{(s-2k-2)}\rangle = (-1)^{k}(2l_{2})^k|\ve^{(s-2)}\rangle~,
\ee
\esubeq
where the states $|\ve^{(s-1)}\rangle$ and $|\ve^{(s-2)}\rangle$ correspond to independent, traceful gauge parameters.

We insert (\ref{Phi Field}) into (\ref{Fronsdal Action Dim Reduced}) and normal order the oscillators to eliminate $b$ and $b^+$, giving
\bea
\label{Action Dim Reduced No External Oscilators}
    \mathcal{L}( \psi^{(s-k)}) &=& \frac{1}{2}\sum_{k=0}^s~\frac{1}{k!}\left\{\langle \psi^{(s-k)}|\left[-\frac{(k+1)(k-2)}{2}\square+l^+_1\Big(l_{2}^+l_{2}+1+\frac{k(k-1)}{4}\Big)l_1\right.\right.\non \\
    &&\left.-l_{2}^+(l_1)^2-(l^+_{1})^2l_{2}-2l_{2}^+\square l_{2}+\frac{(k-1)(k-2)^2}{4}m^2+(k+2)m^2l_{2}^+l_{2}\right]| \psi^{(s-k)}\rangle\non \\
    &&+\langle \psi^{(s-k)}|\left[\frac{(k-1)(k-4)}{4}ml_{1}^++ml_{2}^+l_{1}^+l_{2}+\frac{k-4}{2}ml_{2}^+l_1\right]| \psi^{(s-k-1)}\rangle \non \\
    &&+k\langle \psi^{(s-k)}|\left[\frac{(k-2)(k-5)}{4}ml_{1}+ml_{2}^+l_{1}l_{2}+\frac{k-5}{2}ml_{1}^+l_{2}\right]| \psi^{(s-k+1)}\rangle \non \\
    &&-\langle \psi^{(s-k)}|\left[\Big(\square-\frac{k}{2}m^2\Big)l_{2}^++\frac{1}{2}(l_{1}^+)^2-\frac{1}{2}l_{2}^+l_{1}^+l_{1}\right]| \psi^{(s-k-2)}\rangle \non \\
    &&-k(k-1)\langle \psi^{(s-k)}|\left[\Big(\square-\frac{k-2}{2}m^2\Big)l_{2}+\frac{1}{2}(l_{1})^2-\frac{1}{2}l_{1}^+l_{2}l_{1}\right]| \psi^{(s-k+2)}\rangle \non \\
    &&\left.+\frac{1}{2}\langle \psi^{(s-k)}|ml_{2}^+l_1^+| \psi^{(s-k-3)}\rangle+\frac{k(k-1)(k-2)}{2}\langle \psi^{(s-k)}|ml_{2}l_1| \psi^{(s-k+3)}\rangle\right\}~.
\eea
We then split this into odd and even terms and make use of (\ref{traceful Fields}) to get the Lagrangian in terms of the four independent states as
\bea
\label{Action All Indpendent Fields}
    &&\mathcal{L}\,( \psi^{(s)}, \psi^{(s-1)}, \psi^{(s-2)}, \psi^{(s-3)}) \non \\
    &&~~~= \frac{1}{2}\sum_{k=0}^{[s/2]}~\frac{1}{(2k)!}\left[k\langle \psi^{(s-2)}|+2(k-1)\langle \psi^{(s)}|l_{2}^+\right](2l_{2}^+)^{k-1}\non \\
    &&~~~~~~ \times\bigg\{\bigg[-2k(k-1)\square+l^+_1\Big(\frac{k(2k+1)}{2}-l_{2}^+l_{2}\Big)l_1\non \\
    &&~~~~~~-kl_{2}^+(l_1)^2+(l^+_{1})^2l_{2}+2l_{2}^+\square l_{2}\bigg](2l_{2})^{k-1}| \psi^{(s-2)}\rangle\non \\
    &&~~~~~~+\bigg[-(2k^2-2k+1)\square+l^+_1\Big(\frac{2k^2+k-2}{2}-l_{2}^+l_{2}\Big)l_1-(k-1)l_{2}^+(l_1)^2\non \\
    &&~~~~~~+(l^+_{1})^2l_{2}+2l_{2}^+\square l_{2}+(2k-1)(k-1)m^2-2m^2l_{2}^+l_{2}\bigg](2l_{2})^k| \psi^{(s)}\rangle\non \\
    &&~~~~~~+\left[\frac{k(2k-3)}{2}ml_{1}^+-ml_{2}^+l_{1}^+l_{2}-kml_{2}^+l_1\right](2l_{2})^{k-1}| \psi^{(s-3)}\rangle \non \\
    &&~~~~~~+\left[\frac{(k-2)(2k+1)}{2}ml_{1}^+-ml_{2}^+l_{1}^+l_{2}-(k-2)ml_{2}^+l_1\right](2l_{2})^k| \psi^{(s-1)}\rangle \non \\
    &&~~~~~~+k(k-1)(2k-3)ml_{1}(2l_{2})^{k-2}| \psi^{(s-3)}\rangle+k(k-1)(2k-7)ml_{1}(2l_{2})^{k-1}| \psi^{(s-1)}\rangle \non \\
    &&~~~~~~+k(2k-1)(k-1)(l_{1})^2(2l_{2})^{k-2}| \psi^{(s-2)}\rangle+k(2k-1)(k-2)(l_{1})^2(2l_{2})^{k-1}| \psi^{(s)}\rangle\bigg\} \non \\
    &&~~~~~~+\frac{1}{2}\sum_{k=0}^{[s/2]}~\frac{1}{(2k+1)!}\left[k\langle \psi^{(s-3)}|+2(k-1)\langle \psi^{(s-1)}|l_{2}^+\right](2l_{2}^+)^{k-1}\non \\
    &&~~~~~~ \times\bigg\{\left[-2k^2\square+l^+_1\left(\frac{k(2k+3)}{2}-l_{2}^+l_{2}\right)l_1-kl_{2}^+(l_1)^2+(l^+_{1})^2l_{2}+2l_{2}^+\square l_{2}\right.\non \\
    &&~~~~~~\left.+\frac{k(2k-1)}{2}m^2-m^2l_{2}^+l_{2}\right](2l_{2})^{k-1}| \psi^{(s-3)}\rangle\non \\
    &&~~~~~~+\left[-(2k^2+1)\square+l^+_1\left(\frac{(2k-1)(k+2)}{2}-l_{2}^+l_{2}\right)l_1-(k-1)l_{2}^+(l_1)^2\right.\non \\
    &&~~~~~~\left.+(l^+_{1})^2l_{2}+2l_{2}^+\square l_{2}+\frac{3k(2k-1)}{2}m^2-3m^2l_{2}^+l_{2}\right](2l_{2})^k| \psi^{(s-1)}\rangle\non \\
    &&~~~~~~-\left[\frac{k(2k-1)}{2}ml_{1}^+-ml_{2}^+l_{1}^+l_{2}-\frac{2k+3}{2}ml_{2}^+l_1\right](2l_{2})^k| \psi^{(s-2)}\rangle \non \\
    &&~~~~~~-\left[\frac{2k^2-k-2}{2}ml_{1}^+-ml_{2}^+l_{1}^+l_{2}-\frac{2k-1}{2}ml_{2}^+l_1\right](2l_{2})^{k+1}| \psi^{(s)}\rangle \non \\
    &&~~~~~~-\frac{k(2k+1)(2k-1)}{2}ml_{1}(2l_{2})^{k-1}| \psi^{(s-2)}\rangle-\frac{(k-2)(2k+1)(2k-1)}{2}ml_{1}(2l_{2})^k| \psi^{(s)}\rangle \non \\
    &&~~~~~~+k(2k+1)(k-1)(l_{1})^2(2l_{2})^{k-2}| \psi^{(s-3)}\rangle+k(2k+1)(k-2)(l_{1})^2(2l_{2})^{k-1}| \psi^{(s-1)}\rangle\bigg\}~. \non \\
\eea
We then use (\ref{Gauge Transformation Dim Reduced}) to find the gauge transformations of each independent field as
\bsubeq
\label{Other Independent Gauge Transformations}
\bea
    \d| \psi^{(s)}\rangle &=& \ri l^{+}_1|\ve^{(s-1)}\rangle~, \\
    \non \\
    \d| \psi^{(s-1)}\rangle&=&\ri l_1^+|\ve^{(s-2)}\rangle+\ri m|\ve^{(s-1)}\rangle~,\\
    \non \\
    \d| \psi^{(s-2)}\rangle&=&-2\ri l_1^+l_{2}|\ve^{(s-1)}\rangle+2\ri m|\ve^{(s-2)}\rangle~,\\
    \non \\
    \d| \psi^{(s-3)}\rangle&=&-2\ri l_1^+l_{2}|\ve^{(s-2)}\rangle-6\ri ml_{2}|\ve^{(s-1)}\rangle~.
\eea
\esubeq
We note here that the physical field is contained within $ \psi^{(s)}$ as its traceless part. When deciding our reality conditions, it is appropriate to choose those in which such a field is real, and hence $ \psi^{(s)}$ should be taken to be real. We see from the above gauge conditions that such a requirement imposes reality conditions on the gauge parameters and other fields. By requiring all fields and gauge parameters to be purely real or imaginary we end up with the reality conditions
\bsubeq
\label{Reality Conditions traceful}
\be
    \overline{\j}^{(s-k)}_{\m_1\dots\m_{s-k}} = (-1)^{k} \psi^{(s-k)}_{\m_1\dots\m_{s-k}}~, \qquad k=0,1,2, 3~,
\ee
and
\be
\label{Reality Conditions Traceful Gauge Parameter}
    \overline{\ve}^{(s-k)}_{\m_1\dots\m_{s-k}} = -(-1)^{k}\ve^{(s-k)}_{\m_1\dots\m_{s-k}}~, \qquad k=1,2~.
\ee
\esubeq

It should be noted that the states used here, which coincide\footnote{In the case of \cite{Buchbinder:2015, Buchbinder:2022lsi}, a factor of $\ri$ must be absorbed into the gauge parameters, which simply swaps the sign of the reality conditions (\ref{Reality Conditions Traceful Gauge Parameter}).} with those in \cite{Lindwasser, Buchbinder:2015, Buchbinder:2022lsi}, differ from those in Pashnev's model \cite{Pashnev:1989}, yet they are related through the redefinitions
\bsubeq
\bea
    | \psi^{(s-2)}\rangle\rightarrow -\ri l_{2}| \psi^{(s)}\rangle-\frac{\ri}{2}| \psi^{(s-2)}\rangle~, \\
    \non \\
    | \psi^{(s-3)}\rangle\rightarrow -\ri l_{2}| \psi^{(s-1)}\rangle-\frac{\ri}{2}| \psi^{(s-3)}\rangle~.
\eea
\esubeq
Owing to this, equivalence to either of these models is enough to demonstrate equivalence to the other. The field states in \cite{Asano:2019smc} are obtained from our field states via the redefinitions
\bsubeq
\bea
    |\j^{(s-1)}\rangle\rightarrow -\ri|\j^{(s-1)}\rangle~,\\
    \non \\
    |\j^{(s-2)}\rangle\rightarrow -l_2|\j^{(s)}\rangle+\frac{1}{2}|\j^{(s-2)}\rangle~,\\
    \non \\
    |\j^{(s-3)}\rangle\rightarrow -\ri l_2|\j^{(s-1)}\rangle-\frac{\ri}{6}|\j^{(s-3)}\rangle
\eea
and the gauge parameter must also be redefined according to
\bea
    |\ve^{(s-2)}\rangle\rightarrow \ri|\ve^{(s-2)}\rangle~.
\eea
\esubeq

We will keep the independent states to be those with transformations (\ref{Other Independent Gauge Transformations}) when demonstrating equivalence to the KZ theory as the transformation laws are more suited to our analysis.

\subsection{Equivalence to the Klishevich-Zinoviev theory} \label{Pashnev-KZ}
In the Lagrangian (\ref{Action All Indpendent Fields}), there are
\be
    \binom{d+s-1}{s}+\binom{d+s-2}{s-1}+\binom{d+s-3}{s-2}+\binom{d+s-4}{s-3} =\binom{d+s}{s}-\binom{d+s-4}{s-4}
\ee
degrees of freedom after the reality conditions (\ref{Reality Conditions traceful}) are imposed, which is exactly equal to the degrees of freedom present in the KZ theory. It follows then that there should be some way to recast the model in terms of double-traceless fields similar to the KZ theory. To achieve this, we must decompose the traceful fields into traceless fields and then use these to construct double-traceless fields.

Given a symmetric and traceful tensor field $\j^{(s)}$, we can always decompose it into symmetric and traceless tensor fields $\vf^{(s-2k)}$, $k=0,\dots,[s/2]$, as follows:
\be
\label{Rank k Traceless Decomp}
     \psi^{(s)}_{\m_1\dots\m_s}=\sum^{[s/2]}_{k=0}\eta_{(\m_1\m_2}\dots\eta_{\m_{2k-1}\m_{2k}}\varphi^{(s-2k)}_{\m_{2k+1}\dots\m_{s})}~.
\ee
In oscillator formalism, this is equivalent to the decomposition
\be
\label{Rank k Traceless Decomp Oscilator}
    | \psi^{(s)}\rangle=\sum^{[s/2]}_{k=0}(2l_{2}^+)^k|\varphi^{(s-2k)}\rangle~.
\ee

Before proceeding, we note that for traceless states $|\vf^{(s)}\rangle$ we have
\bsubeq
\be
    l_{2}l_{2}^+|\varphi^{(s)}\rangle = \left(s+\frac{d}{2}\right)|\varphi^{(s)}\rangle~,
\ee
which generalizes to
\be
\label{Trace of Metric Contraction}
    l_{2}(l_{2}^+)^{k}|\varphi^{(s)}\rangle = k\left(s+\frac{d}{2}+k-1\right)(l_{2}^+)^{k-1}|\varphi^{(s)}\rangle~.
\ee
\esubeq

We now decompose each state in (\ref{Other Independent Gauge Transformations}) in terms of traceless states through (\ref{Rank k Traceless Decomp Oscilator}) to get
\bsubeq
\label{Sum Gauge Transformations}
\bea
    \sum_{k=0}^{[s/2]}(2l_{2}^+)^k\d|\vf^{(s-2k)}_0\rangle &=& \sum_{k=0}^{[(s-1)/2]}\ri(2l_{2}^+)^kl_1^+|\x^{(s-2k-1)}\rangle ~,\\
    \non \\
    \sum_{k=0}^{[(s-1)/2]}(2l_{2}^+)^k\d|\vf^{(s-2k-1)}_1\rangle &=& \left(\sum_{k=0}^{[s/2]-1}\ri(2l_{2}^+)^kl_1^+|\x^{(s-2k-2)}\rangle\right)\non \\
    &&+\left(\sum_{k=0}^{[(s-1)/2]}\ri m(2l_{2}^+)^k|\x^{(s-2k-1)}\rangle\right) ~, \\
    \non \\
    \sum_{k=0}^{[s/2]-1}(2l_{2}^+)^k\d|\vf^{(s-2k-2)}_2\rangle &=& -\left(\sum_{k=1}^{[(s-1)/2]}\ri k\left(2s+d-2k-4\right)(2l_{2}^+)^{k-1}l_1^+|\x^{(s-2k-1)}\rangle\right)\non \\
    &&+\left(\sum_{k=0}^{[s/2]-1}2\ri m(2l_{2}^+)^k|\x^{(s-2k-2)}\rangle\right) ~,\\
    \non \\
    \sum_{k=0}^{[(s-1)/2]-1}(2l_{2}^+)^k\d|\vf^{(s-2k-3)}_3\rangle &=& -\left(\sum_{k=1}^{[s/2]-1}\ri k\left(2s+d-2k-6\right)(2l_{2}^+)^{k-1}l_1^+|\x^{(s-2k-2)}\rangle\right)\non \\
    &&-\left(\sum_{k=1}^{[(s-1)/2]}3\ri k\left(2s+d-2k-4\right)m(2l_{2}^+)^{k-1}|\x^{(s-2k-1)}\rangle\right)~, \non\\
\eea
\esubeq
where the field states $|\varphi_j^{(s-2k-j)}\rangle$ are the traceless states contained within $| \psi^{(s-j)}\rangle$ and $|\x^{(s-2k-j)}\rangle$ are the traceless states contained within $|\ve^{(s-j)}\rangle$. Since $l_1^+|\x^{(s-k)}\rangle$ is traceful, it should be decomposed into two traceless fields. We take
\be
    |\l^{(s-k+1)}\rangle=l_1^+|\x^{(s-k)}\rangle-\frac{2}{2s+d-2k-2}l_{2}^+l_1|\x^{(s-k)}\rangle~,
\ee
to be the traceless part of $l_1^+|\x^{(s-k)}\rangle$, hence the decomposition is
\be
    l_1^+|\x^{(s-k)}\rangle = |\l^{(s-k+1)}\rangle+\frac{2}{2s+d-2k-2}l_{2}^+l_1|\x^{(s-k)}\rangle~.
\ee
Inserting this into (\ref{Sum Gauge Transformations}) allows us to isolate the gauge transformation of each traceless field state, leaving
\bsubeq
\label{Traceless Gauge Transformation Oscillator}
\bea
    \d|\vf^{(s-2k)}_0\rangle &=& \ri|\l^{(s-2k)}\rangle+\ri\frac{2}{2s+d-4k}l_1|\x^{(s-2k+1)}\rangle ~,\\
    \non \\
    \d|\vf^{(s-2k-1)}_1\rangle &=& \ri|\l^{(s-2k-1)}\rangle+\ri\frac{2}{2s+d-4k-2}l_1|\x^{(s-2k)}\rangle+\ri m|\x^{(s-2k-1)}\rangle ~, \\
    \non \\
    \d|\vf^{(s-2k-2)}_2\rangle &=& -\ri(k+1)\left(2s+d-2k-6\right)|\l^{(s-2k-2)}\rangle-\ri\frac{2k(2s+d-2k-4)}{2s+d-4k-4}l_1|\x^{(s-2k-1)}\rangle\non \\
    &&+2\ri m|\x^{(s-2k-2)}\rangle ~,\\
    \non \\
    \d|\vf^{(s-2k-3)}_3\rangle &=& -\ri(k+1)\left(2s+d-2k-8\right)|\l^{(s-2k-3)}\rangle-\ri\frac{2k(2s+d-2k-6)}{2s+d-4k-6}l_1|\x^{(s-2k-2)}\rangle\non \\
    &&-3\ri (k+1)\left(2s+d-2k-6\right)m|\x^{(s-2k-3)}\rangle~, \non\\
\eea
\esubeq
where it is understood that
\be
    |\x^{(s+1)}\rangle = |\x^{(s)}\rangle = 0~.
\ee

We then take the field redefinitions
\bsubeq
\label{Redefinitions 1}
\bea
    |\vf^{(s-2k)}_2\rangle\rightarrow \frac{k(2s+d-2k-4)}{2}|\vf^{(s-2k)}_0\rangle+\frac{1}{2}|\vf^{(s-2k)}_2\rangle~,
\eea
and
\bea
    |\vf^{(s-2k-1)}_3\rangle\rightarrow \frac{k(2s+d-2k-6)}{2}|\vf^{(s-2k-1)}_1\rangle+\frac{1}{2}|\vf^{(s-2k-1)}_3\rangle~,
\eea
\esubeq
which leads to the transformations
\bsubeq
\bea
    \d|\vf^{(s-2k)}_2\rangle &=& \ri\,\frac{2s+d-4k-2}{2s+d-4k}l_1|\x^{(s-2k+1)}\rangle+\ri m|\x^{(s-2k)}\rangle ~,\\
    \non \\
    \d|\vf^{(s-2k-1)}_3\rangle &=& \ri\,\frac{2s+d-4k-4}{2s+d-4k-2}l_1|\x^{(s-2k)}\rangle\non \\
    &&-\ri k(2s+d-2k-3)m|\x^{(s-2k-1)}\rangle ~.
\eea
\esubeq
We also take
\bsubeq
\label{Redefinitions 2}
\bea
    |\vf^{(s-2k)}_0\rangle\rightarrow |\vf^{(s-2k)}_0\rangle-\frac{2}{2s+d-4k-2}|\vf^{(s-2k)}_2\rangle~,
\eea
and
\bea
    |\vf^{(s-2k-1)}_1\rangle\rightarrow |\vf^{(s-2k-1)}_1\rangle-\frac{2}{2s+d-4k-4}|\vf^{(s-2k-1)}_3\rangle~,
\eea
\esubeq
which leads to the transformations
\bsubeq
\bea
    \d|\vf^{(s-2k)}_0\rangle &=& \ri|\l^{(s-2k)}\rangle-\ri \,\frac{2}{2s+d-4k-2}m|\x^{(s-2k)}\rangle ~,\\
    \non \\
    \d|\vf^{(s-2k-1)}_1\rangle &=& \ri|\l^{(s-2k-1)}\rangle+\ri \,\frac{(2k+1)(2s+d-2k-4)}{2s+d-4k-4}m|\x^{(s-2k-1)}\rangle ~.
\eea
\esubeq
With these, we construct two sets of double-traceless states
\bsubeq
\label{Redefinitions 3}
\bea
    |\f^{(s-2k)}\rangle &=& |\vf_0^{(s-2k)}\rangle+\frac{2}{2s+d-4k-6}l_{2}^+|\vf^{(s-2k-2)}_2\rangle~, \\
    \non \\
    |\f^{(s-2k-1)}\rangle &=& |\vf_1^{(s-2k-1)}\rangle+\frac{2}{2s+d-4k-8}l_{2}^+|\vf^{(s-2k-3)}_3\rangle~,
\eea
\esubeq
whose gauge transformations are
\bsubeq
\bea
    \d|\f^{(s-2k)}\rangle &=& \ri l_{1}^+|\x^{(s-2k-1)}\rangle-\ri \,\frac{2}{2s+d-4k-2}m|\x^{(s-2k)}\rangle\non \\
    &&+\ri \,\frac{2}{2s+d-4k-6}m\,l_{2}^+|\x^{(s-2k-2)}\rangle~, \\
    \non \\
    \d|\f^{(s-2k-1)}\rangle &=& \ri l_{1}^+|\x^{(s-2k-2)}\rangle+\ri \,\frac{(2k+1)(2s+d-2k-4)}{2s+d-4k-4}m|\x^{(s-2k-1)}\rangle\non \\
    &&-\ri \,\frac{2(k+1)(2s+d-2k-5)}{2s+d-4k-8}m\,l_{2}^+|\x^{(s-2k-3)}\rangle~.
\eea
\esubeq
Now, consider rescalings of the form
\bsubeq
\bea
    |\x^{(s-2k)}\rangle&\rightarrow& \ri a_k|\x^{(s-2k)}\rangle~, \\
    \non \\
    |\f^{(s-2k)}\rangle&\rightarrow& b_k|\f^{(s-2k)}\rangle~, \\
    \non \\
    |\x^{(s-2k-1)}\rangle&\rightarrow&\frac{b_k}{s-2k}|\x^{(s-2k-1)}\rangle~, \\
    \non \\
    |\f^{(s-2k-1)}\rangle&\rightarrow&\ri (s-2k-1)a_{k+1}|\f^{(s-2k-1)}\rangle~,
\eea
\esubeq
for some sets of constants $\{a_k\},\{b_k\}$.
Then the new gauge transformations are
\bsubeq
\label{Gauge Transformation Oscilator Scaling}
\bea
    \d|\f^{(s-2k)}\rangle &=& \frac{\ri}{s-2k}l_{1}^+|\x^{(s-2k-1)}\rangle+\frac{2}{2s+d-4k-2}\frac{a_k}{b_k}\,m|\x^{(s-2k)}\rangle\non \\
    &&-\frac{2}{2s+d-4k-6}\frac{a_{k+1}}{b_k}\,m\,l_{2}^+|\x^{(s-2k-2)}\rangle~, \\
    \non \\
    \d|\f^{(s-2k-1)}\rangle &=& \frac{\ri}{s-2k-1}l_{1}^+|\x^{(s-2k-2)}\rangle+\frac{(2k+1)(2s+d-2k-4)}{(2s+d-4k-4)(s-2k)(s-2k-1)}\frac{b_k}{a_{k+1}}\,m|\x^{(s-2k-1)}\rangle\non \\
    &&-\frac{2(k+1)(2s+d-2k-5)}{(2s+d-4k-8)(s-2k-2)(s-2k-1)}\frac{b_{k+1}}{a_{k+1}}\,m\,l_{2}^+|\x^{(s-2k-3)}\rangle~,
\eea
\esubeq
where all fields and gauge parameters are now real as a consequence of the reality conditions (\ref{Reality Conditions traceful}). Expressing the KZ gauge transformations (\ref{k^th Gauge Transformation}) in the oscillator formalism leads to
\begin{equation}
\label{Zinoviev Gauge Transformation Oscillator Form}
    \begin{split}
        \delta|\phi^{(k)}\rangle =~ & \frac{\ri}{k} l_1^+|\xi^{(k-1)}\rangle+\alpha_k|\xi^{(k)}\rangle-\beta_kl_{2}^+|\xi^{(k-2)}\rangle ~.
    \end{split}
\end{equation}
Comparing these with (\ref{Gauge Transformation Oscilator Scaling}), we see that for the theories to have equivalent gauge transformations we require
\bsubeq
\bea
    \a_{s-2k}=\frac{2}{2s+d-4k-2}\frac{a_k}{b_k}\,m~, \\
    \non \\
    \b_{s-2k}=\frac{2}{2s+d-4k-6}\frac{a_{k+1}}{b_k}\,m~, \\
    \non \\
    \a_{s-2k-1}=\,\frac{(2k+1)(2s+d-2k-4)}{(2s+d-4k-4)(s-2k)(s-2k-1)}\frac{b_k}{a_{k+1}}\,m~, \\
    \non \\
    \b_{s-2k-1}=\frac{2(k+1)(2s+d-2k-5)}{(2s+d-4k-8)(s-2k-2)(s-2k-1)}\frac{b_{k+1}}{a_{k+1}}\,m~.
\eea
\esubeq
The latter two equations will hold given the first two equations and are therefore redundant. Combining what remains leads to the recursion relation
\bea
    a_{k+1}=\frac{2\a_{s-2k-1}}{\a_{s-2k}(2s+d-4k-2)(s-2k-1)}a_k~,
\eea
solved by
\bsubeq
\label{KZ to Pashnev Parameters}
\bea
    a_{k+1}=\left(\prod_{n=1}^{k}\frac{2\a_{s-2n-1}}{\a_{s-2n}(2s+d-4n-2)(s-2n-1)}\right)a_1~,
\eea
where $a_1$ can be freely chosen. The condition on $b_k$ is then
\be
    b_k = \frac{2m}{\a_{s-2n}(2s+d-4k-2)}\left(\prod_{n=1}^{k-1}\frac{2\a_{s-2n-1}}{\a_{s-2n}(2s+d-4n-2)(s-2n-1)}\right)a_1~.
\ee
\esubeq
Thus, the field transformations of the dimensional reduction are equivalent to those of the KZ theory by a field redefinition. It follows then that the two models are equivalent and hence the KZ theory is equivalent to the Pashnev theory. With this relation now understood, we now move to discuss how the gauge may be fixed in the Pashnev theory to reproduce the model presented in section \ref{Singh Hagen Correspondence}.


\subsection{Gauge fixing to $d$-dimensional Singh-Hagen theory} \label{Pashnev Gauge Fixing}
It has been pointed out in \cite{Lindwasser, Buchbinder:2015, Buchbinder:2022lsi} that by making use of the gauge freedom described by the rank $s-1$ and $s-2$ traceful gauge parameters in the quartet theory, 
 one can simply eliminate the rank $s-1$ and $s-2$ fields, that is one can impose the gauge conditions
\bea
|\j^{(s-1)}\rangle=0~, \qquad |\j^{(s-2)}\rangle =0~.
\label{4.55}
\eea
The remaining fields will be of the same tensor types as in the Singh-Hagen theory.
It has further been claimed in \cite{Buchbinder:2022lsi} that the resulting action ``will coincide with the Singh-Hagen action.'' This claim is not quite correct. 

Let us recall our construction of the $d$-dimensional Singh-Hagen model given 
in Section \ref{Singh Hagen Correspondence}.
 We worked with the KZ theory and completely fixed the corresponding gauge freedom 
 imposing the gauge conditions \eqref{gauge_condition}, that is  by
 enforcing all double-traceless fields to have vanishing traceless parts, except for the field of rank $s$. 
 The resulting gauge-fixed action coincided with the Singh-Hagen action in the $d=4$ case. 
Our gauge conditions can be recast in the framework of the Pashnev theory.
 By working backwards from (\ref{Redefinitions 3}) to (\ref{Redefinitions 2}) we find that the gauge conditions 
 \eqref{gauge_condition} are equivalent to
\bsubeq
\bea
    |\vf^{(s-k)}_0\rangle-\frac{2}{2s+d-2k-2}|\vf^{(s-k)}_2\rangle=0~, \qquad 1\leq k\leq s~,
\eea
and
\bea
    |\vf^{(s-k-1)}_1\rangle-\frac{2}{2s+d-2k-4}|\vf^{(s-k-1)}_3\rangle=0~, \qquad 0\leq k\leq s-1~.
\eea
\esubeq
With  these gauge fixing conditions, the Pashnev theory leads to the $d$-dimensional Singh-Hagen model.
From (\ref{d-dimSH}), we see that the kinetic term for the field $\varphi^{(s-3)}$ is proportional to
\bea
    \varphi^{(s-3)\mu_{1}\dots\mu_{s-3}}\left(\square-\frac{3(d+2s-6)^2}{2(d+2s-8)(d+2s-7)}m^2\right)\varphi^{(s-3)}_{\mu_{1}\dots\mu_{s-3}}~,
 \label{4.57}
\eea
which for $d=4$ is
\bea
    \varphi^{(s-3)\mu_{1}\dots\mu_{s-3}}\left(\square-\frac{3(s-1)^2}{(s-2)(2s-3)}m^2\right)\varphi^{(s-3)}_{\mu_{1}\dots\mu_{s-3}}~. \label{4.58}
\eea
Now, let us suppose that we have imposed the alternative gauge conditions \eqref{4.55}. Then it may be shown that the kinetic term for  the field $  \varphi^{(s-3)}$ is proportional to
\bea
    \varphi^{(s-3)\m_1\dots\m_{s-3}}\left(\square-\frac{1}{4}m^2\right)\varphi^{(s-3)}_{\m_1\dots\m_{s-3}}~, \label{4.59}
\eea
which is a result that is independent of the value of $d$. 
This clearly differs from \eqref{4.58}. We also note that these fields only appear for $s\geq3$, for which \eqref{4.57} will never coincide with \eqref{4.59}. We see then that different gauge conditions lead to different actions for the massive spin-$s$ fields. 

It was pointed out by Lindwasser \cite{Lindwasser} that the conditions \eqref{4.55} do not completely fix the gauge freedom.\footnote{Lindwasser analysed the residual gauge freedom on the mass shell and demonstrated that this invariance  can be completely fixed by gauging away certain on-shell fields, ensuring that the remaining field satisfied the Fierz-Pauli equations (\ref{Fierz Pauli}). A similar approach for slightly different gauge conditions was carried out earlier by Pashnev \cite{Pashnev:1989}.}
In Appendix \ref{AppendixB} it will be demonstrated that the gauged-fixed action, which is  obtained by imposing \eqref{4.55}, does not reproduce all the equations of motion which correspond to the original gauge-invariant action in conjunction with the Noether identities. Therefore this functional cannot be identified with a $d$-dimensional the Singh-Hagen action, unlike the action constructed 
in Section \ref{Singh Hagen Correspondence}.


\section{Conclusion} \label{Section5}

In this paper we have demonstrated that the KZ theory \cite{Klishevich:1997pd} is equivalent to the Pashnev theory \cite{Pashnev:1989} upon an appropriate identification of the field variables. Similar considerations may be used to show that, in the integer-spin case, the gauge-invariant models derived in \cite{Buchbinder:2008ss, Asano:2019smc, Lindwasser} are also equivalent to the Pashnev theory. The latter theory remains largely unknown to (or unappreciated within) the higher-spin community, in spite of its conceptual significance. 

We believe that the approach advocated by Pashnev can be used to derive gauge-invariant formulations for off-shell massive higher-spin supermultiplets in three dimensions by applying dimensional reduction to the off-shell massless supersymmetric higher-spin models in four dimensions \cite{KPS,KS}. Attempts of generalizing the Zinoviev formulation 
\cite{Zinoviev:2001} for such a goal have so far failed.\footnote{In four dimensions, an off-shell $\cN=1$ supersymmetric generalisation of the Singh-Hagen models was derived in \cite{Koutrolikos:2020}.}

We have also derived a $d$-dimensional extension of the Singh-Hagen model \cite{Singh:1974} through fixing the gauge freedom of the KZ theory. By using the equivalence between the KZ theory and the Pashnev theory, we found how such a choice of gauge manifests itself in the latter. Our analysis revealed that enforcing the gauge conditions \eqref{4.55} off-shell introduced in \cite{Lindwasser, Buchbinder:2015, Buchbinder:2022lsi} does not reproduce the Singh-Hagen action \cite{Singh:1974} in the $d=4$ case, nor do they lead to a $d$-dimensional extension of this action. The resulting gauge-fixed action will also not reproduce all of the equations of motion 
in the KZ theory in the gauge \eqref{4.55}. Some of these equations must be added as a constraint when dealing with this gauge-fixed action.
 Therefore, the gauge-fixed action obtained by enforcing the gauge conditions \eqref{4.55} cannot be a $d$-dimensional extension of the Singh-Hagen action.
These issues will be further discussed 
 in Appendix \ref{AppendixB}.
\\

\section*{Acknowledgments}
We acknowledge the referee for raising an important issue that led to the inclusion of Appendix B.
We are grateful to Emmanouil Raptakis and Mirian Tsulaia for useful comments on the manuscript. We also thank I. L. Buchbinder, Maxim Grigoriev, Vladimir Krykhtin, Lukas Lindwasser, Ruslan Metsaev, Alexander Reshetnyak, Eugene Skvortsov,  and Mirian Tsulaia for email correspondence and bringing important references to our attention. The work of SMK is supported in part by the Australian Research Council, project No. DP230101629. The work of AJF is supported by the Australian Government Research Training Program.

\renewcommand{\theequation}{\Alph{section}.\arabic{equation}}

\renewcommand{\thesubsection}{\arabic{subsection}.}

\appendix
\section{Quantization of the Klishevich-Zinoviev model~~~} \label{AppendixA}
Covariant BRST quantization of the Zinoviev theory in (A)dS$_d$ \cite{Zinoviev:2001}
has been studied by Metsaev in \cite{Metsaev:2008fs, Metsaev:2014vda}. 
In this appendix we carry out the Faddeev-Popov quantization of the KZ theory in Minkowski space ${\mathbb M}^d$. Unlike in \cite{Metsaev:2008fs, Metsaev:2014vda}, we will not use the oscillator formalism throughout the quantization process. However, the final results will be recast in terms of this formalism.

This appendix is based on the results reported in our earlier unpublished work \cite{Fegebank:2023}. Analogous results have been obtained in recent publications \cite{Cangemi:2022bew, Cangemi:2023ysz, Cangemi:2023bpe}.

The dynamical variables of the spin-$s$ KZ theory are symmetric double-traceless (for $k\leq 4$) gauge fields 
 $\f^{(k)} $, with $k = s, s-1, \dots , 0$. 
In the $k>0$ case, associated with  $\f^{(k)} $
is the symmetric and traceless gauge parameter $\x^{(k-1)}$.
Let $\overline{ c}^{ (k-1 )} $ and $ c^{(k-1)}$ be the symmetric and traceless Faddeev-Popov ghosts
corresponding to  $\x^{(k-1)}$.
For path integrals, 
 the following shorthand is adopted
\bsubeq
\begin{align}
    \int\mathcal{D}(\phi;k) &:= \int\prod^{k}_{j=0}\mathcal{D}\phi^{(j)}~,\\
    \noalign{\vspace{10pt}}\int\mathcal{D}(\phi, c\,;k) &:= \int\Big(\prod^{k}_{j=1}\mathcal{D}\phi^{(j)}\mathcal{D}\overline{ c}^{ (j-1 )}\mathcal{D} c^{(j-1)}\Big)\mathcal{D}\phi^{(0)}~.
\end{align}
\esubeq
In order to evaluate the ghost contributions, the following identity is noted:
Faddeev-Popov determinants are realized in terms of the path integral according to the general rule:
\begin{subequations} \label{Determinant of M} 
\begin{align}
&       \det M^{(k)}
 = \int \mathcal{D}\overline{ c}^{(k)}\mathcal{D} c^{(k)}\non \\
       &\qquad \qquad  \times \exp\left\{-\ri \int \rd^dx \int \rd^dx'\,\overline{ c}^{(k)\mu_1\dots\mu_k}(x)
       M_{\mu_1\dots \mu_{k}}{}^{\n_1 \dots \n_k} (x,x') \, c^{(k)}_{\nu_{1}\dots\nu_k}(x')\right\}~. 
\end{align}
Here  $M^{(k)}$ is an operator acting on the space of symmetric and traceless fields $\j^{(k)}$, 
\bea
M^{(k)}:  c^{(k)}_{\mu_{1}\dots\mu_k}(x) \to
\int \rd^dx'\, M_{\mu_1\dots \mu_{k}}{}^{\n_1 \dots \n_k} (x,x')\,  c^{(k)}_{\nu_{1}\dots\nu_k}(x')~.
\eea
\end{subequations}


\subsection{Spin $s=2$} \label{Spin 2}
The KZ theory \eqref{Full Action for spin s} 
is an irreducible gauge theory 
 (following the terminology of the Batalin-Vilkovisky formalism \cite{BV})
and  can be quantized \`a la Faddeev and Popov \cite{Faddeev:1967fc}.
Before considering its quantization 
in the general $s\geq 2$ case, 
it is worth investigating how the process is carried out for the simplest $s=2$ and $s=3$ values. 

In the spin-2 case, the action (\ref{Full Action for spin s}) is
\begin{align}
        S  = & \int \rd^dx \,\left\{-\frac{1}{2}\partial^{\mu}\phi^{(2)\nu\lambda}\partial_\mu\phi^{(2)}_{\nu\lambda}+\partial_\mu\phi^{(2)\mu\lambda}\partial^\nu\phi^{(2)}_{\nu\lambda}+\frac{1}{2}\partial^\mu\widetilde{\phi}^{(2)}\partial_\mu\widetilde{\phi}^{(2)}\right.\non \\
        & \hspace{40pt}+\partial^\mu\partial^\nu\phi^{(2)}_{\mu\nu}\widetilde{\phi}^{(2)}-\frac{1}{2}\partial^\mu\phi^{(1)\nu}\partial_\mu\phi^{(1)}_\nu+\frac{1}{2}\partial^\mu\phi^{(1)}_\mu\partial^\nu\phi^{(1)}_\nu\non \\
        & \hspace{40pt}-\frac{1}{2}\partial^\mu\phi^{(0)}\partial_\mu\phi^{(0)}+a_2\phi^{(1)\nu}\partial^\mu\phi^{(2)}_{\mu\nu}+b_2\widetilde{\phi}^{(2)}\partial^\mu\phi^{(1)}_\mu\label{Spin 2 Action}\\
        & \hspace{40pt}+d_2\phi^{(2)\mu\nu}\phi^{(2)}_{\mu\nu}+e_2(\widetilde{\phi}^{(2)})^2+a_1\phi^{(0)}\partial^\mu\phi^{(1)}_\mu\non \\
        & \hspace{40pt}\left.-f_2\widetilde{\phi}^{(2)}\phi^{(0)}+d_1\phi^{(1)\mu}\phi^{(1)}_\mu+d_0(\phi^{(0)})^2\right\}~,\non 
\end{align}
and the gauge transformation (\ref{k^th Gauge Transformation}) reads
\bsubeq
    \label{Spin 2 Gauge Transformation}
    \begin{align}
        \delta\phi^{(2)}_{\mu\nu} & = \frac{1}{2}\left(\partial_\mu\xi^{(1)}_\nu+\partial_\nu\xi^{(1)}_\mu\right)-\beta_2\eta_{\mu\nu}\xi^{(0)}~,\label{spin 2 gauge transformation 1}\\
        \noalign{\vspace{10pt}} \delta\phi^{(1)}_\mu & = \partial_\mu\xi^{(0)}+\alpha_1\xi^{(1)}_\mu~,\label{spin 2 gauge transformation 2}\\
        \noalign{\vspace{10pt}} \delta\phi^{(0)} & = \alpha_0\xi^{(0)}~.\label{spin 2 gauge transformation 3}
    \end{align} 
\esubeq

To carry out the Faddeev-Popov scheme, suitable gauge conditions are required. We choose the following gauge-fixing functions:
\bsubeq
    \begin{align}
        &\Xi^{(1)}_{\mu}
        -\chi^{(1)}_{\mu} 
        =2\partial^\nu\phi^{(2)}_{\mu\nu}-\partial_\mu\widetilde{\phi}^{(2)}-m\sqrt{2}\phi^{(1)}_\mu-\chi^{(1)}_\mu ~,\label{Spin 2 Gauge Fixing 3}\\
        \noalign{\vspace{10pt}} &\Xi^{(0)}
        -\chi^{(0)} 
        = \partial^\mu\phi^{(1)}_\mu
        -\frac{m}{\sqrt{2}}\widetilde{\phi}^{(2)}
        -\frac{2d-2}{d-2}\frac{m^2}{\alpha_0}\phi^{(0)}-\chi^{(0)} \label{Spin 2 Gauge Fixing 4}~,
    \end{align}
\esubeq
where $\chi^{(1)}$ and $\chi^{(0)}$ are background fields. The explicit expressions for $\Xi^{(1)}$ and $\Xi^{(0)}$  have been chosen so that their gauge variations are
\bsubeq
    \begin{align}
        \delta\Xi^{(1)}_\mu=(\square-m^2)\xi^{(1)}_\mu \label{Xi 1 Varied}~,\\
        \noalign{\vspace{10pt}} \delta\Xi^{(0)}=(\square-m^2)\xi^{(0)}\label{Xi 0 Varied}~.
    \end{align}
\esubeq
With these, we have the partition function 
\begin{equation}
    Z^{(2)} = \int\mathcal{D}(\phi;2)\Delta^{(1)}\Delta^{(0)}\delta\left[\Xi^{(1)}_\mu-\chi^{(1)}_\mu\right]\delta\left[\Xi^{(0)}-\chi^{(0)}\right]\re^{\ri S} \label{Spin 2 Partition Function 1}~,
\end{equation}
where the objects $\Delta^{(1)}$ and $\Delta^{(0)}$ are 
the Faddeev-Popov determinants, 
\bsubeq
    \begin{align}
        \Delta^{(0)} &= \det \left(\frac{\delta \Xi^{(0)}(x)}{\delta\xi^{(0)}(x')}\right) = \det \left[(\square-m^2)\delta^d(x-x')\right]~,\label{Spin 2 First Delta 1}\\
        \noalign{\vspace{10pt}} \Delta^{(1)} &= \det\left( \frac{\delta \Xi^{(1)}_\mu(x)}{\d \xi^{(1)}_\nu(x')}\right) = \det[ \d_\m{}^\n(\square-m^2)\delta^d(x-x')]~. \label{Spin 2 Second Delta 1}
    \end{align}
\esubeq

Since the partition function \eqref{Spin 2 Partition Function 1} is independent of the background fields $\chi^{(1)}$ and $\chi^{(0)}$, we can average over them with a convenient weight of the form
\bea
    \exp\left\{-\frac{\ri}{2}\int \rd^dx \,\bigg[\frac{\chi^{(1)\mu}\chi^{(1)}_\mu}{\omega_1}+\frac{(\chi^{(0)})^2}{\omega_0}\bigg]\right\}~,
\eea
where $\omega_1$ and $\omega_0$ are constants that will be chosen in such a way as to diagonalize the action as well as cause all divergence terms to vanish. 
Doing so, (\ref{Spin 2 Partition Function 1}) becomes
\begin{align}
        Z^{(2)} = & \int\mathcal{D}(\phi;2)\Delta^{(1)}\Delta^{(0)}\non \\
        & \times\exp\left\{-\frac{\ri}{2}\int \rd^dx \,\bigg[\frac{1}{\omega_1}\big(2\partial^\nu\phi^{(2)}_{\mu\nu}-\partial_\mu\widetilde{\phi}^{(2)}-m\sqrt{2}\phi^{(1)}_\mu\big)^2\right.
        \label{Spin 2 Partition Function 2}\\
        & \hspace{100pt}\left.+\frac{1}{\omega_0}\Big(\partial^\mu\phi^{(1)}_\mu
        -\frac{m}{\sqrt{2}}\widetilde{\phi}^{(2)}
        -\frac{2d-2}{d-2}\frac{m^2}{\alpha_0}\phi^{(0)}\Big)^2\bigg]\right\}\re^{\ri S} ~.\non
\end{align}
Substituting the classical action (\ref{Spin 2 Action}) in \eqref{Spin 2 Partition Function 2} yields
\begin{align}
        Z^{(2)} = & \int\mathcal{D}(\phi;2)\Delta^{(1)}\Delta^{(0)}\exp\left\{\ri\int \rd^dx \,\bigg[-\frac{1}{2}\partial^{\mu}\phi^{(2)\nu\lambda}\partial_\mu\phi^{(2)}_{\nu\lambda}+\Big(1-\frac{2}{\omega_1}\Big)\partial_\mu\phi^{(2)\mu\lambda}\partial^\nu\phi^{(2)}_{\nu\lambda}\right.\non \\
        & +\frac{1}{2}\Big(1-\frac{1}{\omega_1}\Big)\partial^\mu\widetilde{\phi}^{(2)}\partial_\mu\widetilde{\phi}^{(2)}+\Big(1-\frac{2}{\omega_1}\Big)\partial^\mu\partial^\nu\phi^{(2)}_{\mu\nu}\widetilde{\phi}^{(2)}-\frac{1}{2}\partial^\mu\phi^{(1)\nu}\partial_\mu\phi^{(1)}_\nu\non \\
        & +\Big(d_1-\frac{m^2}{\omega_1}\Big)\phi^{(1)\mu}\phi^{(1)}_\mu+\frac{1}{2}\Big(1-\frac{1}{\omega_0}\Big)\partial^\mu\phi^{(1)}_\mu\partial^\nu\phi^{(1)}_\nu-\frac{1}{2}\partial^\mu\phi^{(0)}\partial_\mu\phi^{(0)}\non \\
        & +\Big(a_2+\frac{2\sqrt{2}m}{\omega_1}\Big)\phi^{(1)\nu}\partial^\mu\phi^{(2)}_{\mu\nu}+\Big(b_2+\frac{m}{\sqrt{2}\omega_0}+\frac{\sqrt{2}m}{\omega_1}\Big)\widetilde{\phi}^{(2)}\partial^\mu\phi^{(1)}_\mu\label{Spin 2 Partition Function 3}\\
        & +d_2\phi^{(2)\mu\nu}\phi^{(2)}_{\mu\nu}+\Big(e_2-\frac{m^2}{4\omega_0}\Big)(\widetilde{\phi}^{(2)})^2-\Big(f_2+\frac{\sqrt{2}(d-1)}{d-2}\frac{m^3}{\alpha_0\omega_0}\Big)\widetilde{\phi}^{(2)}\phi^{(0)}\non \\
        & \left.+\Big(a_1+\frac{2d-2}{d-2}\frac{m^2}{\alpha_0\omega_0}\Big)\phi^{(0)}\partial^\mu\phi^{(1)}_\mu+\Big(d_0-\frac{(2d-2)^2}{2(d-2)^2}\frac{m^4}{(\alpha_0)^2\omega_0}\Big)(\phi^{(0)})^2\bigg]\right\} \non ~.
\end{align}
To get our desired result, we choose 
\bea
    \omega_1 = 2~, \qquad
    \omega_0 = 1~.
\eea
With such a choice, 
making use of (\ref{Zinoviev Coefficients}) and (\ref{Alpha k}) gives
\begin{align}
        Z^{(2)} = & \int\mathcal{D}(\phi;2)\Delta^{(1)}\Delta^{(0)}\exp\left\{\ri\int \rd^dx \,\bigg[\frac{1}{2}\phi^{(2)\mu\nu}(\square-m^2)\phi^{(2)}_{\mu\nu}-\frac{1}{4}\widetilde{\phi}^{(2)}(\square-m^2)\widetilde{\phi}^{(2)}\right.\non \\
        & \left.+\frac{1}{2}\phi^{(1)\mu}(\square-m^2)\phi^{(1)}_\mu+\frac{1}{2}\phi^{(0)}(\square-m^2)\phi^{(0)}\bigg]\right\} \label{Spin 2 Partition Function 6}~.
 \end{align}

It only remains to recast 
the ghost contributions in terms of path integrals.
In accordance with (\ref{Determinant of M}) we rewrite (\ref{Spin 2 First Delta 1}) and (\ref{Spin 2 Second Delta 1}) as follows:
\begin{subequations}
\bea
    \Delta^{(0)} &=& \int \mathcal{D}\overline{ c}^{(0)}\mathcal{D} c^{(0)}\exp\left\{-\ri\int \rd^dx \,\:\overline{ c}^{(0)}(\square-m^2) c^{(0)}\right\} \label{Spin 2 First Delta 2}~,\\
    \Delta^{(1)} &=& \int \mathcal{D}\overline{ c}^{(1)}\mathcal{D} c^{(1)}\exp\left\{-\ri\int \rd^dx \,\:\overline{ c}^{(1)\mu}(\square-m^2) c^{(1)}_\mu\right\}\label{Spin 2 Second Delta 2}~.
\eea
\end{subequations}

With these, (\ref{Spin 2 Partition Function 6}) becomes
\begin{align}
        Z^{(2)} = & \int\mathcal{D}(\phi, c\,;2)\exp\left\{\ri\int \rd^dx \,\bigg[\frac{1}{2}\phi^{(2)\mu\nu}(\square-m^2)\phi^{(2)}_{\mu\nu}-\frac{1}{4}\widetilde{\phi}^{(2)}(\square-m^2)\widetilde{\phi}^{(2)}\right.\non\\
        & -\overline{ c}^{(1)\mu}(\square-m^2) c^{(1)}_\mu+\frac{1}{2}\phi^{(1)\mu}(\square-m^2)\phi^{(1)}_\mu-\overline{ c}^{(0)}(\square-m^2) c^{(0)}\label{Spin 2 Partition Function 7}\\
        & \left.+\frac{1}{2}\phi^{(0)}(\square-m^2)\phi^{(0)}\bigg]\right\} ~,\non
\end{align}
which is the fully diagonalized partition function for the spin-2 case.


\subsection{Spin $s=3$} \label{Spin 3}
A similar procedure is carried out for the spin-3 field. The corresponding classical action is
\begin{align}
        S  = & \int \rd^dx \,\bigg\{-\frac{1}{2}\partial^\mu\phi^{(3)\nu\lambda\rho}\partial_\mu\phi^{(3)}_{\nu\lambda\rho}+\frac{3}{2}\partial_\mu\phi^{(3)\mu\lambda\rho}\partial^\nu\phi^{(3)}_{\nu\lambda\rho}+\frac{3}{2}\partial^\mu\widetilde{\phi}^{(3)\nu}\partial_\mu\widetilde{\phi}^{(3)}_\nu\non \\
        & \hspace{40pt} +3\partial_\mu\partial_\nu\phi^{(3)\mu\nu\lambda}\widetilde{\phi}^{(3)}_{\lambda}+\frac{3}{4}\partial_\mu\widetilde{\phi}^{(3)\mu}\partial_\nu\widetilde{\phi}^{(3)\nu}+a_3\phi^{(2)\nu\lambda}\partial^\mu\phi^{(3)}_{\mu\nu\lambda}\non \\
        & \hspace{40pt} +b_3m\widetilde{\phi}^{(3)\nu}\partial^\mu\phi^{(2)}_{\mu\nu}+c_3\partial_\mu\widetilde{\phi}^{(3)\mu}\widetilde{\phi}^{(2)}+d_3\phi^{(3)\mu\nu\lambda}\phi^{(3)}_{\mu\nu\lambda}+e_3\widetilde{\phi}^{(3)\mu}\widetilde{\phi}^{(3)}_{\mu}\non \\
        & \hspace{40pt} -f_3\widetilde{\phi}^{(3)\mu}\phi^{(1)}_{\mu}-\frac{1}{2}\partial^{\mu}\phi^{(2)\nu\lambda}\partial_\mu\phi^{(2)}_{\nu\lambda}+\partial_\mu\phi^{(2)\mu\lambda}\partial^\nu\phi^{(2)}_{\nu\lambda}
        \label{Spin 3 Action}\\
        & \hspace{40pt} +\frac{1}{2}\partial^\mu\widetilde{\phi}^{(2)}\partial_\mu\widetilde{\phi}^{(2)}+\partial^\mu\partial^\nu\phi^{(2)}_{\mu\nu}\widetilde{\phi}^{(2)}-\frac{1}{2}\partial^\mu\phi^{(1)\nu}\partial_\mu\phi^{(1)}_\nu\non \\
        & \hspace{40pt} +\frac{1}{2}\partial^\mu\phi^{(1)}_\mu\partial_\nu\phi^{(1)}_\nu-\frac{1}{2}\partial^\mu\phi^{(0)}\partial_\mu\phi^{(0)}+a_2\phi^{(1)\nu}\partial^\mu\phi^{(2)}_{\mu\nu}+b_2\widetilde{\phi}^{(2)}\partial^\mu\phi^{(1)}_\mu\non \\
        & \hspace{40pt} +d_2\phi^{(2)\mu\nu}\phi^{(2)}_{\mu\nu}+e_2(\widetilde{\phi}^{(2)})^2-f_2\widetilde{\phi}^{(2)}\phi^{(0)}+a_1m\phi^{(0)}\partial^\mu\phi^{(1)}_\mu+d_1\phi^{(1)\mu}\phi^{(1)}_\mu\non \\
        & \hspace{40pt} +d_0(\phi^{(0)})^2\bigg\} ~.
        \non 
 \end{align}
It is invariant under the gauge transformations
\bsubeq
\label{Spin 3 Gauge Transformations}
    \begin{align}
        \delta\phi^{(3)}_{\mu\nu\lambda} &= \frac{1}{3}\big(\partial_\mu\xi^{(2)}_{\nu\lambda}+\partial_\nu\xi^{(2)}_{\mu\lambda}+\partial_\lambda\xi^{(2)}_{\mu\nu}\big)-\beta_3\big(\eta_{\mu\nu}\xi^{(1)}_{\lambda}+\eta_{\lambda\mu}\xi^{(1)}_{\nu}+\eta_{\nu\lambda}\xi^{(1)}_{\mu}\big)~, \label{Spin 3 Gauge Transformation 1}\\
        \noalign{\vspace{10pt}}\delta\phi^{(2)}_{\mu\nu} &= \frac{1}{2}\big(\partial_\mu\xi^{(1)}_\nu+\partial_\nu\xi^{(1)}_\mu\big)+\alpha_2\xi^{(2)}_{\mu\nu}-\beta_2\eta_{\mu\nu}\xi^{(0)}~, \label{Spin 3 Gauge Transformation 2}\\
        \noalign{\vspace{10pt}} \delta\phi^{(1)}_\mu &= \partial_\mu\xi^{(0)}+\alpha_1\xi^{(1)}_\mu~, \label{Spin 3 Gauge Transformation 3}\\
        \noalign{\vspace{10pt}} \delta\phi^{(0)} &= \alpha_0\xi^{(0)}~. \label{Spin 3 Gauge Transformation 4}
    \end{align} 
\esubeq

To quantize the theory,  we choose  the following gauge-fixing  functions:
\bsubeq \label{5.3}
    \begin{align}
        &\Xi^{(2)}_{\mu\nu}-\chi^{(2)}_{\mu\nu} = 3\partial^\lambda\phi^{(3)}_{\mu\nu\lambda} - 3\partial_{(\mu}\widetilde{\phi}^{(3)}_{\nu)} - \sqrt{3}m\phi^{(2)}_{\mu\nu} + \frac{\sqrt{3}}{d}\eta_{\mu\nu}m\widetilde{\phi}^{(2)}-\chi^{(2)}_{\mu\nu} ~,\label{Spin 3 Gauge Fixing 7}\\
        \noalign{\vspace{10pt}} &\Xi^{(1)}_{\mu}-\chi^{(1)}_{\mu} =  2\partial^\nu\phi^{(2)}_{\mu\nu}
        -\sqrt{3}m\widetilde{\phi}^{(3)}_\mu  - \partial_\mu\widetilde{\phi}^{(2)} - 2m\sqrt{\frac{d+1}{d}}\phi^{(1)}_\mu-\chi^{(1)}_\mu ~,\label{Spin 3 Gauge Fixing 8}\\
        \noalign{\vspace{10pt}} &\Xi^{(0)}-\chi^{(0)} = \partial^\mu\phi^{(1)}_\mu  -m\sqrt{\frac{d+1}{d}}\widetilde{\phi}^{(2)} - m\sqrt{\frac{3d}{d-2}}\phi^{(0)}-\chi^{(0)}~,\label{Spin 3 Gauge Fixing 9}
    \end{align}
\esubeq
where $\chi^{(2)} $, $\chi^{(1)}$ and $\chi^{(0)} $ are background fields. Here both  $\Xi^{(2)}$  and $\c^{(2)}$ are symmetric and traceless. 
The gauge-fixing functions \eqref{5.3} have been chosen so that $\Xi^{(2)}$ 
varies as
\bea
    \delta\Xi^{(2)} _{\mu\nu} = (\square-m^2)\xi^{(2)}_{\mu\nu}~,
\eea
while the gauge variations $ \d \Xi^{(1)}$ and $\d \Xi^{(0)}$ are given by 
(\ref{Xi 1 Varied}) and (\ref{Xi 0 Varied}), respectively.
The partition function is then
\begin{equation}
    Z^{(3)} = \int\mathcal{D}(\phi;3)\Delta^{(2)}\Delta^{(1)}\Delta^{(0)}\delta\big[\Xi^{(2)}_{\mu\nu}-\chi^{(2)}_{\mu\nu}\big]\delta\big[\Xi^{(1)}_\mu-\chi^{(1)}_\mu\big]\delta\big[\Xi^{(0)}-\chi^{(0)}\big]\re^{\ri S}~, \label{Spin 3 Partition Function 1}
\end{equation}
where once again $\Delta^{(0)}$ and $\Delta^{(1)}$ are ghost contributions of spin 0 and 1 while $\Delta^{(2)}$ is the ghost contribution of spin 2. 

Since the partition function \eqref{Spin 3 Partition Function 1} is independent of the background fields $\chi^{(2)}$, $\chi^{(1)}$ and $\chi^{(0)}$, we can average over them with a convenient weight of the form
\bea
    \exp\left\{-\frac{\ri}{2}\int \rd^dx \,\bigg[\frac{\chi^{(2)\mu\nu}\chi^{(2)}_{\mu\nu}}{\omega_2}+\frac{\chi^{(1)\mu}\chi^{(1)}_\mu}{\omega_1}+\frac{(\chi^{(0)})^2}{\omega_0}\bigg]\right\}~.
    \label{5.6}
\eea
Upon doing so, the relation (\ref{Spin 3 Partition Function 1}) turns into
\bea
        Z^{(3)} &= & \int\mathcal{D}(\phi;3)\Delta^{(2)}\Delta^{(1)}\Delta^{(0)}\displaystyle \exp
        \bigg\{-\frac{\ri}{2}\int \rd^dx \,\non \\
        && \times
        \bigg[
        \frac{1} {\omega_2} \Big(3\partial_\lambda\phi^{(3)\mu\nu\lambda} - 3\partial^{(\mu}\widetilde{\phi}^{(3)\nu)} - \sqrt{3}m\phi^{(2)\mu\nu} + \frac{\sqrt{3}}{d}\eta^{\mu\nu}m\widetilde{\phi}^{(2)}\Big)^2 \non \\
        && +\frac{1}{\omega_1}  \Big(2\partial^\nu\phi^{(2)}_{\mu\nu} 
        -\sqrt{3}m\widetilde{\phi}^{(3)}_\mu 
        - \partial_\mu\widetilde{\phi}^{(2)} 
        - 2m\sqrt{\frac{d+1}{d}}\phi^{(1)}_\mu\Big)^2 
        \label{Spin 3 Partition Function 2}\\
        && 
        +\frac{1} {\omega_0} \Big( \partial^\mu\phi^{(1)}_\mu
        -m\sqrt{\frac{d+1}{d}}\widetilde{\phi}^{(2)}  - m\sqrt{\frac{3d}{d-2}}\phi^{(0)}\Big)^2  \bigg]\bigg\}\re^{\ri S}\non ~.
\eea
This is then combined with (\ref{Spin 3 Action}) to get
\bea
        Z^{(3)} &= & \int\mathcal{D}(\phi;3)\Delta^{(2)}\Delta^{(1)}\Delta^{(0)}\displaystyle \exp\left\{\ri\int \rd^dx \,\bigg[-\frac{1}{2}\partial^\mu\phi^{(3)\nu\lambda\rho}\partial_\mu\phi^{(3)}_{\nu\lambda\rho}
        +\Big(\frac{3}{2}-\frac{9}{2\omega_2}\Big)
        \partial_\mu\phi^{(3)\mu\lambda\rho}\partial^\nu\phi^{(3)}_{\nu\lambda\rho}\right.\non \\
        && +\Big(\frac{3}{2}-\frac{9}{4\omega_2}\Big)\partial^\mu\widetilde{\phi}^{(3)\nu}\partial_\mu\widetilde{\phi}^{(3)}_\nu+\Big(3-\frac{9}{\omega_2}\Big)\partial_\mu\partial_\nu\phi^{(3)\mu\nu\lambda}\widetilde{\phi}^{(3)}_{\lambda}+\Big(\frac{3}{4}-\frac{9}{4\omega_2}\Big)\partial_\mu\widetilde{\phi}^{(3)\mu}\partial_\nu\widetilde{\phi}^{(3)\nu}\non \\
        && +\Big(a_3+\frac{3\sqrt{3}m}{\omega_2}\Big)\phi^{(2)\nu\lambda}\partial^\mu\phi^{(3)}_{\mu\nu\lambda}+\Big(b_3+\frac{2\sqrt{3}m}{\omega_1}+\frac{3\sqrt{3}m}{\omega_2}\Big)\widetilde{\phi}^{(3)\nu}\partial^\mu\phi^{(2)}_{\mu\nu}\non \\
        && +\Big(c_3+\frac{\sqrt{3}m}{\omega_1}\Big)\partial_\mu\widetilde{\phi}^{(3)\mu}\widetilde{\phi}^{(2)}+\frac{m^2}{2}\phi^{(3)\mu\nu\lambda}\phi^{(3)}_{\mu\nu\lambda}+\Big(e_3-\frac{3m^2}{2\omega_1}\Big)\widetilde{\phi}^{(3)\mu}\widetilde{\phi}^{(3)}_{\mu}\non \\
        && -\Big(f_3+\frac{2m^2}{\omega_1}\sqrt{\frac{3(d+1)}{d}}\Big)\widetilde{\phi}^{(3)\mu}\phi^{(1)}_{\mu}-\frac{1}{2}\partial^{\mu}\phi^{(2)\nu\lambda}\partial_\mu\phi^{(2)}_{\nu\lambda}+\Big(1-\frac{2}{\omega_1}\Big)\partial_\mu\phi^{(2)\mu\lambda}\partial^\nu\phi^{(2)}_{\nu\lambda}\non \\
        && +\Big(\frac{1}{2}-\frac{1}{2\omega_1}\Big)\partial^\mu\widetilde{\phi}^{(2)}\partial_\mu\widetilde{\phi}^{(2)}+\Big(1-\frac{2}{\omega_2}\Big)\partial^\mu\partial^\nu\phi^{(2)}_{\mu\nu}\widetilde{\phi}^{(2)}-\frac{1}{2}\partial^\mu\phi^{(1)\nu}\partial_\mu\phi^{(1)}_\nu\label{Spin 3 Partition Function 3}\\
        && +\Big(\frac{1}{2}-\frac{1}{2\omega_0}\Big)\partial^\mu\phi^{(1)}_\mu\partial_\nu\phi^{(1)}_\nu+\frac{1}{2}\partial^\mu\phi^{(0)}\partial_\mu\phi^{(0)}+\Big(a_2+\frac{4m}{\omega_1}\sqrt{\frac{d+1}{d}}\Big)\phi^{(1)\nu}\partial^\mu\phi^{(2)}_{\mu\nu}\non \\
        && +\Big(b_2+\frac{m}{\omega_0}\sqrt{\frac{d+1}{d}}+\frac{2m}{\omega_1}\sqrt{\frac{d+1}{d}}\Big)\widetilde{\phi}^{(2)}\partial^\mu\phi^{(1)}_\mu-\frac{3}{2\omega_2}m^2\phi^{(2)\mu\nu}\phi^{(2)}_{\mu\nu}\non \\
        && +\Big(e_2-\frac{3m^2}{2d\omega_2}-\frac{(d+1)m^2}{2d\omega_0}\Big)(\widetilde{\phi}^{(2)})^2-\Big(f_2+\frac{m^2}{\omega_0}\sqrt{\frac{3(d+1)}{d-2}}\Big)m^2\widetilde{\phi}^{(2)}\phi^{(0)}\non \\
        && +\Big(a_1+\frac{m}{\omega_0}\sqrt{\frac{3d}{d-2}}\Big)\phi^{(0)}\partial^\mu\phi^{(1)}_\mu+\Big(d_1-\frac{2(d+1)m^2}{d\omega_1}\Big)\phi^{(1)\mu}\phi^{(1)}_\mu\non \\
        && \left.+\Big(d_0-\frac{3dm^2}{2(d-2)\omega_0}\Big)(\phi^{(0)})^2\bigg]\right\}~.\non 
\eea
It may be seen that the coefficients in \eqref{5.6} that diagonalize the action and remove all terms with divergences (such as $\partial_\mu\phi^{(3)\mu\lambda\rho}\partial^\nu\phi^{(3)}_{\nu\lambda\rho}$) are:
\bea
    \omega_2 = 3~, \qquad \omega_1 = 2~, \qquad \omega_0 = 1~.
\eea
Taking into account the relations (\ref{Zinoviev Coefficients}) and integrating by parts, one arrives at
\bea
        Z^{(3)} &= & \int\mathcal{D}(\phi;3)\Delta^{(2)}\Delta^{(1)}\Delta^{(0)}\displaystyle \exp\left\{\ri\int \rd^dx \,\bigg[\frac{1}{2}\phi^{(3)\mu\nu\lambda}(\square-m^2)\phi^{(3)}_{\mu\nu\lambda}\right.\non\\
        && -\frac{3}{4}\widetilde{\phi}^{(3)\mu}(\square-m^2)\widetilde{\phi}^{(3)}_{\mu}+\frac{1}{2}\phi^{(2)\mu\nu}(\square-m^2)\phi^{(2)}_{\mu\nu}-\frac{1}{4}\widetilde{\phi}^{(2)}(\square-m^2)\widetilde{\phi}^{(2)}\label{Spin 3 Partition Function 4}\\
        && \left.+\frac{1}{2}\phi^{(1)\mu}(\square-m^2)\phi^{(1)}_{\mu}+\frac{1}{2}\phi^{(0)}(\square-m^2)\phi^{(0)}\bigg]\right\}~.\non
\eea

Finally, it only remains to massage 
the ghost contributions. 
The spin-0 and spin-1 ghost contributions are just (\ref{Spin 2 First Delta 2}) and (\ref{Spin 2 Second Delta 2}) as before. The spin-3 contribution is given by
\bea
    \Delta^{(2)} = \det\Bigg(\frac{\delta \Xi^{(2)}_{\mu\nu} (x)}{\delta \xi^{(2)}_{\lambda\rho}(x')}\Bigg) = \det \bigg[ \hat{\d}_{\m \n}{}^{\l \r} (\square-m^2) \d^d(x-x')
      \bigg] ~, \label{Spin 3 Third Delta 1}
\eea
where $ \hat{\d}_{\m \n}{}^{\l \r} $ is the Kronecker delta on the space of symmetric traceless second-rank tensors,
\bea
  \hat{\d}_{\m \n}{}^{\l \r} = \d_{(\m}{}^\l \d_{\n)}{}^\r - \frac{1}{d} \eta_{\m\n} \eta^{\l\r}~.
  \eea
Upon making use of (\ref{Determinant of M}), one obtains
\begin{equation}
    \Delta^{(2)} = \int \mathcal{D}\overline{ c}^{(2)}\mathcal{D} c^{(2)}\exp\left\{-\ri\int \rd^dx \,\:\overline{ c}^{(2)\mu\nu}(\square-m^2) c^{(2)}_{\mu\nu}\right\} \label{Third Delta 2}~.
\end{equation}
With this, the partition (\ref{Spin 3 Partition Function 4}) takes the form
\bea
        Z^{(3)} &= & \int\mathcal{D}(\phi, c\,;3)\displaystyle \exp\left\{\ri\int \rd^dx \,\bigg[\frac{1}{2}\phi^{(3)\mu\nu\lambda}(\square-m^2)\phi^{(3)}_{\mu\nu\lambda}-\frac{3}{4}\widetilde{\phi}^{(3)\mu}(\square-m^2)\widetilde{\phi}^{(3)}_{\mu}\right.\non\\
        && -\overline{ c}^{(2)\mu\nu}(\square-m^2) c^{(2)}_{\mu\nu}+\frac{1}{2}\phi^{(2)\mu\nu}(\square-m^2)\phi^{(2)}_{\mu\nu}-\frac{1}{4}\widetilde{\phi}^{(2)}(\square-m^2)\widetilde{\phi}^{(2)}\label{Spin 3 Partition Function 5}\\
        && -\overline{ c}^{(1)\mu}(\square-m^2) c^{(1)}_\mu+\frac{1}{2}\phi^{(1)\mu}(\square-m^2)\phi^{(1)}_{\mu}-\overline{ c}^{(0)}(\square-m^2) c^{(0)}\non\\
        && \left.+\frac{1}{2}\phi^{(0)}(\square-m^2)\phi^{(0)}\bigg]\right\}~.\non
  \eea
It is seen that the gauge-fixed action is fully diagonalized.


\subsection{Arbitrary integer spin $s$} \label{Spin s}
In the spin-$s$ case, the gauge-invariant action has the form (\ref{Full Action for spin s}). The corresponding  gauge freedom  is given by the transformations (\ref{k^th Gauge Transformation}) where $0\leq k \leq s$. 

To quantize the theory, we introduce the following symmetric and traceless gauge-fixing functions:  
\begin{align}
        \Xi^{(k)}_{\mu_1\dots\mu_k}-\chi^{(k)}_{\mu_1\dots\mu_k} &= (k+1)\partial^\mu\phi^{(k+1)}_{\mu\mu_1\dots\mu_{k}}
        -\frac{(k+2)(k+1)}{2}\alpha_{k+1}\widetilde{\phi}^{(k+2)}_{\mu_1\dots\mu_{k}}
        ~.\non\\
        &\hspace{15pt}-\frac{k(k+1)}{2}\partial_{(\mu_1}\widetilde{\phi}^{(k+1)}_{\mu_2\dots\mu_{k})}-\frac{(k+1)k(d+2k-4)}{2}\beta_{k+1}\phi^{(k)}_{\mu_1\dots\mu_{k}} \non\\
        &\hspace{15pt}+\frac{(k+1)k^2(k-1)}{4}\beta_{k+1}\eta_{(\mu_1\mu_2}\widetilde{\phi}^{(k)}_{\mu_3\dots\mu_{k})}-\chi^{(k)}_{\mu_1\dots\mu_k}
                \label{General kth Gauge Fixing 5}~,
 \end{align}
where $\chi^{(k)}$ is a background symmetric traceless field. 
The gauge-fixing function $\Xi^{(k)}$  has been chosen such that its variation is 
\be
    \delta\Xi^{(k)}_{\mu_1\dots\mu_k} = (\square-m^2)\xi^{(k)}_{\mu_1\dots\mu_k}~.\label{General kth Gauge Fixing 2}
\ee

The partition function is given by 
\begin{equation}
    Z^{(s)} = \int\mathcal{D}(\phi;s)\Big(\prod_{k=0}^{s-1}\Delta^{(k)}\delta\left[\Xi^{(k)}-\chi^{(k)}\right]\Big)\re^{\ri S}~. \label{Spin s Partition Function 1}
\end{equation}
Since the partition function  is independent of the background fields $\chi^{(k)}$, we can average over them with a convenient weight of the form
\bea
    \prod^{s-1}_{k=0}\exp\left\{-\frac{\ri}{2}\int \rd^dx \,\bigg[\frac{\chi^{(k)\mu_1\dots\mu_k}\chi^{(k)}_{\mu_1\dots\mu_k}}{\omega_k}\bigg]\right\}~,
\eea
where $\omega_k$ are constants chosen such that they diagonalize the action. This gives
\begin{equation}
    Z^{(s)} =  \int\mathcal{D}(\phi;s)\left(\prod_{k=0}^{s-1}\Delta^{(k)}\exp\left\{-\frac{\ri}{2}\int \rd^dx \,\bigg[\frac{\Xi^{(k)\mu_1\dots\mu_k}\Xi^{(k)}_{\mu_1\dots\mu_k}}{\omega_k}\bigg]\right\}\right)\re^{\ri S}~.\label{Spin s Partition Function 2}
\end{equation}

We find that $\Xi^{(k)\mu_1\dots\mu_k}\Xi^{(k)}_{\mu_1\dots\mu_k}$ is expanded fully as
\bea
        &&\frac{(k+1)^2(k+2)^2}{4}(\alpha_{k+1})^2\widetilde{\phi}^{(k+2)\mu_1\dots\mu_k}\widetilde{\phi}^{(k+2)}_{\mu_1\dots\mu_k}+(k+1)^2\partial_\mu\phi^{(k+1)\mu\mu_1\dots\mu_k}\partial^\nu\phi^{(k+1)}_{\nu\mu_1\dots\mu_k}\non \\
        &&+\frac{k(k+1)^2}{4}\partial^\mu\widetilde{\phi}^{(k+1)\mu_2\dots\mu_k}\partial_\mu\widetilde{\phi}^{(k+1)}_{\mu_2\dots\mu_k}+\frac{k(k+1)^2(k-1)}{4}\partial^\mu\widetilde{\phi}^{(k+1)\nu\mu_3\dots\mu_k}\partial_\nu\widetilde{\phi}^{(k+1)}_{\mu\mu_3\dots\mu_k}\non \\
        &&+\frac{(k+1)^2k^2(d+2k-4)^2}{4}(\beta_{k+1})^2\phi^{(k)\mu_1\dots\mu_k}\phi^{(k)}_{\mu_1\dots\mu_k}\non \\
        &&-\frac{(k+1)^2k^3(k-1)(d+2k-4)}{8}(\beta_{k+1})^2\widetilde{\phi}^{(k)\mu_3\dots\mu_k}\widetilde{\phi}^{(k)}_{\mu_3\dots\mu_k}\label{Square 2 Right}\\
        &&-(k+2)(k+1)^2\alpha_{k+1}\widetilde{\phi}^{(k+2)\mu_1\dots\mu_k}\partial^\mu\phi^{(k+1)}_{\mu\mu_1\dots\mu_k}+\frac{(k+2)(k+1)^2k}{2}\alpha_{k+1}\widetilde{\phi}^{(k+2)\mu_1\dots\mu_k}\partial_{\mu_1}\widetilde{\phi}^{(k+1)}_{\mu_2\dots\mu_k}\non \\
        &&+\frac{(k+2)(k+1)^2k(d+2k-4)}{2}\alpha_{k+1}\beta_{k+1}\widetilde{\phi}^{(k+2)\mu_1\dots\mu_k}\phi^{(k)}_{\mu_1\dots\mu_k}\non \\
        &&+k(k+1)^2\partial_\mu\phi^{(k+1)\mu\mu_1\dots\mu_k}\partial_{\mu_1}\widetilde{\phi}^{(k+1)}_{\mu_2\dots\mu_k}+(k+1)^2k(d+2k-4)\beta_{k+1}\partial_\mu\phi^{(k+1)\mu\mu_1\dots\mu_k}\phi^{(k)}_{\mu_1\dots\mu_k}\non \\
        &&+\frac{(k+1)^2k^2(d+2k-4)}{2}\beta_{k+1}\partial^{\mu_1}\widetilde{\phi}^{(k+1)\mu_2\dots\mu_k}\phi^{(k)}_{\mu_1\dots\mu_k}~.\non 
 \eea
These terms can be split into massless and massive parts, allowing us to look at how they modify the massive and massless Lagrangian contributions separately. Denoting the new contributions with a prime, the effect of (\ref{Square 2 Right}) on (\ref{Massless Lagrangian Contribution}) is
\bea
        \mathcal{L}'_0(\phi^{(k)})  &= & -\frac{1}{2}\partial^\mu\phi^{(k)\mu_1\dots\mu_k}\partial_\mu\phi^{(k)}_{\mu_1\dots\mu_k}+\frac{k}{2}\partial_\mu\phi^{(k)\mu\mu_2\dots\mu_k}\partial^\nu\phi^{(k)}_{\nu\mu_2\dots\mu_k}\non\\
        && +\frac{k(k-1)}{4}\partial^\mu\widetilde{\phi}^{(k)\mu_3\dots\mu_k}\partial_\mu\widetilde{\phi}^{(k)}_{\mu_3\dots\mu_k}+\frac{k(k-1)}{2}\partial_\mu\partial_\nu\phi^{(k)\mu\nu\mu_3\dots\mu_k}\widetilde{\phi}^{(k)}_{\mu_3\dots\mu_k}\non\\
        && +\frac{k(k-1)(k-2)}{8}\partial_\mu\widetilde{\phi}^{(k)\mu\mu_4\dots\mu_k}\partial^\nu\widetilde{\phi}^{(k)}_{\nu\mu_4\dots\mu_k}+\frac{(k-1)k^2}{2\omega_{k-1}}\partial_\mu\phi^{(k)\mu\nu\mu_3\dots\mu_k}\partial_\nu\widetilde{\phi}^{(k)}_{\mu_3\dots\mu_k} \non\\
        && -\frac{(k-1)k^2(k-2)}{8\omega_{k-1}}\partial^\mu\widetilde{\phi}^{(k)\nu\mu_4\dots\mu_k}\partial_\nu\widetilde{\phi}^{(k)}_{\mu\mu_4\dots\mu_k}-\frac{k^2}{2\omega_{k-1}}\partial_\mu\phi^{(k)\mu\mu_2\dots\mu_k}\partial^\nu\phi^{(k)}_{\nu\mu_2\dots\mu_k}\non\\
        && -\frac{(k-1)k^2}{8\omega_{k-1}}\partial^\mu\widetilde{\phi}^{(k)\mu_3\dots\mu_k}\partial_\mu\widetilde{\phi}^{(k)}_{\mu_3\dots\mu_k}~.\label{Massless Lagrangian Contribution Prime}
\eea
To cancel out all off diagonal and divergence contributions, we choose
\begin{equation}
    \omega_k = k+1~. \label{Omega}
\end{equation}
With such a choice, (\ref{Massless Lagrangian Contribution Prime}) becomes
\bea
        \mathcal{L}'_0(\phi^{(k)})  &= & -\frac{1}{2}\partial^\mu\phi^{(k)\mu_1\dots\mu_k}\partial_\mu\phi^{(k)}_{\mu_1\dots\mu_k}+\frac{k(k-1)}{8}\partial^\mu\widetilde{\phi}^{(k)\mu_3\dots\mu_k}\partial_\mu\widetilde{\phi}^{(k)}_{\mu_3\dots\mu_k}\non\\
        && +\frac{k(k-1)}{2}\partial_\mu\partial_\nu\phi^{(k)\mu\nu\mu_3\dots\mu_k}\widetilde{\phi}^{(k)}_{\mu_3\dots\mu_k}+\frac{(k-1)k}{2}\partial_\mu\phi^{(k)\mu\nu\mu_3\dots\mu_k}\partial_\nu\widetilde{\phi}^{(k)}_{\mu_3\dots\mu_k}\non\\
        && +\frac{k(k-1)(k-2)}{8}\partial_\mu\widetilde{\phi}^{(k)\mu\mu_4\dots\mu_k}\partial^\nu\widetilde{\phi}^{(k)}_{\nu\mu_4\dots\mu_k}\label{Massless Lagrangian Contribution Prime 2}\\
        && -\frac{(k-1)k(k-2)}{8}\partial^\mu\widetilde{\phi}^{(k)\nu\mu_4\dots\mu_k}\partial_\nu\widetilde{\phi}^{(k)}_{\mu\mu_4\dots\mu_k}~,\non
  \eea
and therefore
\begin{equation}
    \begin{split}
        \int \rd^d x\mathcal{L}'_0(\phi^{(k)})  = & \int \rd^dx\bigg[\frac{1}{2}\phi^{(k)\mu_1\dots\mu_k}\square\phi^{(k)}_{\mu_1\dots\mu_k}-\frac{k(k-1)}{8}\widetilde{\phi}^{(k)\mu_3\dots\mu_k}\square\widetilde{\phi}^{(k)}_{\mu_3\dots\mu_k}\bigg]~.
    \end{split}\label{Integrated Massless Lagrangian Contribution Prime}
\end{equation}
Now we check if this value of $\o_k$ diagonalizes the massive Lagrangian contributions. Indeed, by combining the massive terms in (\ref{Square 2 Right}) with (\ref{k^th Massive Lagrangian Contribution}) and (\ref{Omega}), as well as substituting (\ref{Zinoviev Coefficients}) and (\ref{Alpha k}), we have
\be
        \mathcal{L}'_m(\phi^{(k)})  =  -\frac{1}{2}m^2\phi^{(k)\mu_1\dots\mu_k}\phi^{(k)}_{\mu_1\dots\mu_k}+\frac{k(k-1)}{8}m^2\widetilde{\phi}^{(k)\mu_3\dots\mu_k}\widetilde{\phi}^{(k)}_{\mu_3\dots\mu_k}~.
   \ee
Thus, the entire action is diagonalized by choosing (\ref{Omega}). With this, (\ref{Spin s Partition Function 2}) becomes
\bea
        Z^{(s)} &= & \int\mathcal{D}(\phi;s)\Big(\prod_{k=0}^{s-1}\Delta^{(k)}\Big)\non\\
        && \times\exp\Bigg\{\sum^{s}_{k=0}\bigg[\:\ri \int \rd^dx\Big(\frac{1}{2}\phi^{(k)\mu_1\dots\mu_k}(\square-m^2)\phi^{(k)}_{\mu_1\dots\mu_k}\label{Spin s Partition Function 3}\\
        && \hspace{120pt}-\frac{k(k-1)}{8}\widetilde{\phi}^{(k)\mu_3\dots\mu_k}(\square-m^2)\widetilde{\phi}^{(k)}_{\mu_3\dots\mu_k}\Big)\bigg]\Bigg\}~.\non
  \eea

The Faddeev-Popov determinants  $\Delta^{(k)}$ are given by
\bea
    \Delta^{(k)} = \det(\frac{\delta \Xi^{(k)}_{\mu_1\dots\mu_k} (x)}{\delta \xi^{(k)}_{\nu_1\dots\nu_k}(x')}) = \det \bigg[ \hat{\d}_{\m_1  \dots \m_k}{}^{\n_1 \dots \n_k}  (\square-m^2) \d^d(x-x')
      \bigg]~,
\eea
where $\hat{\d}_{\m_1  \dots \m_k}{}^{\n_1 \dots \n_k} $ denotes the Kronecker delta on the space of symmetric traceless rank-$k$ tensors.
Making use of (\ref{Determinant of M}) gives
\begin{equation}
    \Delta^{(k)} = \int\mathcal{D}\overline{ c}^{(k)}\mathcal{D} c^{(k)}\exp\left\{-\ri \int \rd^dx \,\overline{ c}^{(k)\mu_1\dots\mu_k}(\square-m^2) c^{(k)}_{\mu_1\dots\mu_k}\right\}~. \label{k^th Ghost Contribution}
\end{equation}

The full partition function is obtained by inserting (\ref{k^th Ghost Contribution}) in (\ref{Spin s Partition Function 3}), yielding
\bea
        Z^{(s)} &= & \int\mathcal{D}(\phi, c\,;s)\exp\bigg\{\:\ri\int \rd^dx\,\bigg[
        \frac{1}{2}\phi^{(s)\mu_1\dots\mu_s}(\square-m^2)\phi^{(s)}_{\mu_1\dots\mu_s}
              \non\\
        && 
              -\frac{s(s-1)}{8}\widetilde{\phi}^{(s)\mu_3\dots\mu_s}(\square-m^2)\widetilde{\phi}^{(s)}_{\mu_3\dots\mu_s}
               \label{Spin s Partition Function 4}\\
        && +   
             \sum^{s-1}_{k=0}
             \Big(\frac{1}{2}\phi^{(k)\mu_1\dots\mu_k}(\square-m^2)\phi^{(k)}_{\mu_1\dots\mu_k}
               -\frac{k(k-1)}{8}\widetilde{\phi}^{(k)\mu_3\dots\mu_k}(\square-m^2)\widetilde{\phi}^{(k)}_{\mu_3\dots\mu_k}
              \non\\
        &&
               -\overline{ c}^{(k)\mu_1\dots\mu_k}(\square-m^2) c^{(k)}_{\mu_1\dots\mu_k}\Big)\bigg]
        \bigg\}~.\non
    \eea
The structure of \eqref{Spin s Partition Function 4} is similar to the effective action for massive antisymmetric tensor field models in $d$ dimensions \cite{BKP,KTurner}.
It is obvious from  \eqref{Spin s Partition Function 4} that the massive propagators are remarkably simple in the Feynman-like  gauge, which we have used. An alternative quantization scheme is to make use of the unitary gauge 
\eqref{unitary}, in which the gauge freedom is absent and the theory is described by the $d$-dimensional Singh-Hagen model \eqref{d-dimSH}. The corresponding propagators for the massive integer-spin fields in four dimensions
were derived in \cite{Singh:1981aw}.

Making use of (\ref{Spin s Partition Function 4}), we can count 
the number of degrees of freedom, $n(d,s)$, using the relation $Z^{(s)} = \det^{-n(d,s)/2} \big( \Box - m^2\big)$.
A symmetric rank-$k$ tensor in $d$ dimensions has
\bea
    \binom{d+k-1}{k}
    \label{6.15}
\eea
independent components. In the case of a traceless symmetric tensor,  this should be reduced by the number of independent components of a symmetric rank-$(k-2)$ tensor. Given a double-traceless symmetric tensor, 
\eqref{6.15} should be reduced by the number of of a symmetric rank-$(k-4)$ tensor. 
The total degrees of freedom are counted as
\be
    \sum^s_{k=0}\binom{d+k-1}{k}-\sum^{s-4}_{k=0}\binom{d+k-1}{k}-2\left(\sum^{s-1}_{k=0}\binom{d+k-1}{k}-\sum^{s-3}_{k=0}\binom{d+k-1}{k}\right)~,
\ee
which, after some algebra, simplifies to $n(d,s)$ given by  eq. \eqref{3.11}.

All fields in \eqref{Spin s Partition Function 4} have one and the same kinetic operator, $(\Box -m^2)$. 
This degeneracy is characteristic of ${\mathbb M}^d$. In the case of an (anti-)de Sitter background, the structure of kinetic operators depend on the rank of a field, and the partition function becomes a function on the spacetime curvature
\cite{Metsaev:2008fs, Metsaev:2014vda}. 


\subsection{Oscillator form of the partition function} \label{Partition Oscillator}
We see from (\ref{Spin s Partition Function 4}) that the massive field propagators are similar to the $(d+1)$-dimensional massless propagators of the Fronsdal theory. In this sense, the KZ theory ensures that the fields inherit the simple propagators of the Fronsdal theory, which is owed to having the massless parts of the Lagrangian (\ref{Massless Lagrangian Contribution}) to be taken as the Fronsdal Lagrangian. We will now use (\ref{Spin s Partition Function 4}) to construct a partition function in terms of four traceful field states. First we will choose to define our field ket-states\footnote{We here exclude the factor of $\frac{1}{k!}$ to simplify coefficients.}
\be
    |\f^{(k)}\rangle: = \f^{(k)\m_1\dots\m_k}a^+_{\m_1}\dots a^+_{\m_k}|0\rangle~,
\ee
as well as ghost states
\be
    | c^{(k)}\rangle: =  c^{(k)\m_1\dots\m_k}a^+_{\m_1}\dots a^+_{\m_k}|0\rangle~, \qquad | \overline{c}^{(k)}\rangle: =  \overline{c}^{(k)\m_1\dots\m_k}a^+_{\m_1}\dots a^+_{\m_k}|0\rangle~.
\ee
The fields $\f^{(k)}$ are those in the KZ theory and are thus taken to be double traceless and real. The ghost fields $c^{(k)}$ and $\overline{c}^{(k)}$ are taken as traceless and real. This allows us to write (\ref{Spin s Partition Function 4}) as
\bea
    Z^{(s)} &= & \int\mathcal{D}(\phi, c\,;s)\exp\bigg\{\:\ri\int \rd^dx\,\bigg[
    \frac{1}{2}\langle\f^{(s)}|(\square-m^2)(1-s(s-1)l_2^+l_2)|\f^{(s)}\rangle
    \non \\
    && +   
    \sum^{s-1}_{k=0}
    \Big(\frac{1}{2}\langle\f^{(k)}|(\square-m^2)(1-k(k-1)l_2^+l_2)|\f^{(k)}\rangle
    -\langle \overline{c}^{(k)}|(\square-m^2)| c^{(k)}\rangle\Big)\bigg]
    \bigg\}~.
\eea
Before continuing, we note that one could use \eqref{Redefinitions 1}, \eqref{Redefinitions 2} and \eqref{Redefinitions 3} to express these in terms of the quartet fields identified in Section \ref{Dim Reduction}, but doing so will obviously result in the inclusion of off-diagonal terms in the action. The propagators will thus take a more complicated form\footnote{The propagators for such a quartet of fields have been found in \cite{Lindwasser}.}, and hence be more difficult to work with.

By taking the field redefinition
\bea
    |\f^{(k)}\rangle\rightarrow(1-c_kl_2^+l_2)|\f^{(k)}\rangle~, \qquad k\geq 2~,
\eea
where\footnote{Another set of redefinitions where the sign of the square root in $c_k$ is taken as positive will lead to the same result.}
\be
    c_k = \frac{2-\sqrt{4-2k(k-1)(2k+d-4)}}{2k+d-4}~,
\ee
the partition function becomes
\bea
    Z^{(s)} &= & \int\mathcal{D}(\phi, c\,;s)\exp\bigg\{\:\ri\int \rd^dx\,\bigg[
    \frac{1}{2}\langle\f^{(s)}|(\square-m^2)|\f^{(s)}\rangle
    \non \\
    && +   
    \sum^{s-1}_{k=0}
    \Big(\frac{1}{2}\langle\f^{(k)}|(\square-m^2)|\f^{(k)}\rangle
    -\langle \overline{c}^{(k)}|(\square-m^2)| c^{(k)}\rangle\Big)\bigg]
    \bigg\}~,
\eea
and hence the propagators simplify further\footnote{Propagators of this form have been found in \cite{Metsaev:2014vda}, though for a more general background of constant curvature.}. Note that the fields for $k>1$ will now be complex under the reality conditions (\ref{Reality Conditions traceful}), due to the sign under the square root in $c_k$ being negative. We then collect the double-traceless field states into 4 traceful fields through
\bea
    |\j^{(s-j)}\rangle&:=&\sum^{[(s-j)/4]}_{k=0}(2l_2^+)^{2k}|\f^{(s-4k-j)}\rangle~, \qquad j=0,\dots, 3~,
\eea
and
\be
    |C^{(s-j)}\rangle := \sum^{[(s-j)/2]}_{k=0}(2l_2^+)^{k}| c^{(s-2k-j)}\rangle~, \qquad |\overline{C}^{(s-j)}\rangle := \sum^{[(s-j)/2]}_{k=0}(2l_2^+)^{k}| \overline{c}^{(s-2k-j)}\rangle~, \qquad j=1,2~,
\ee
which leads us to
\bea
    Z^{(s)} &= & \int\mathcal{D}(\j,C\,;s)\exp\bigg\{\:\ri\int \rd^dx\,\bigg[
    \sum^{3}_{j=0}\frac{1}{2}\langle\j^{(s-j)}|(\square-m^2)|\j^{(s-j)}\rangle
    \non \\
    && -   
    \sum^{1}_{j=0}
    \langle \overline{C}^{(s-j)}|(\square-m^2)|C^{(s-j)}\rangle\bigg]
    \bigg\}~,
\eea
as the partition function in terms of a quartet of traceful fields.


\section{Alternative gauge fixing conditions}
 \label{AppendixB}

In Section \ref{Singh Hagen Correspondence} we constructed the $d$-dimensional extension of the Singh-Hagen model. This was achieved by starting from the KZ theory and then fixing the corresponding gauge freedom by imposing the gauge conditions \eqref{gauge_condition}.
Alternative gauge fixing conditions \eqref{4.55} have also been discussed in the literature \cite{Lindwasser, Buchbinder:2015, Buchbinder:2022lsi}. Unlike our gauge conditions, these do not completely fix the gauge freedom as pointed out by Lindwasser \cite{Lindwasser}.
Indeed, it follows from \eqref{Other Independent Gauge Transformations} that the  conditions \eqref{4.55} are preserved by residual gauge transformations of the form 
\bea
 l_1^+|\ve^{(s-2)}\rangle+  m|\ve^{(s-1)}\rangle =0~, \qquad
l_1^+l_{2}|\ve^{(s-1)}\rangle- m|\ve^{(s-2)}\rangle=0~.
\eea
As explained in Section  \ref{Pashnev Gauge Fixing},
the gauge conditions \eqref{4.55}  lead to a different expression for the gauge-fixed action, to be denoted $S_{\rm alt}$, than the action $S_{\rm SH}$ corresponding to \eqref{d-dimSH}.
In this appendix we wish to investigate whether the equations of motion for $S_{\rm alt}$ describe dynamics equivalent to that in the Pashnev theory. To make this technical issue clearer, let us first consider two simple examples.

Our first example is
the Stueckelberg model \eqref{1.4}. 
The equations of motion for the action $\widetilde S$ corresponding to the Lagrangian $\widetilde \cL$, 
eq. \eqref{1.4}, are
\bea
    \frac{\d \widetilde{S}}{\d A^\m} = (\square-m^2)A_\m-\pa_\m\pa^\n A_\n+m\pa_\m\vf =0~,\qquad \frac{\d \widetilde{S}}{\d \vf} = -m\pa^\m A_\m+\square\vf =0~.
\eea
The gauge invariance of $\widetilde S$, eq. \eqref{Stueckelberg Gauge Transformations}, is equivalent to the Noether identity
\bea
\label{Stueckelberg Noether}
    -\pa^\m\frac{\d \widetilde{S}}{\d A^\m}+m\frac{\d \widetilde{S}}{\d \vf}\equiv0~.
\eea
Let $S$ be the action corresponding to the Lagrangian $\cL$, eq. \eqref{1.3}. This action is obtained from
 $\widetilde S$ by imposing  
the gauge condition $\vf=0$. The corresponding equation of motion is 
\bea
 \frac{\d S}{\d A_\m}=
 \frac{\d \widetilde{S}}{\d A_\m}\bigg|_{\vf=0}=0~.
 \label{massive_spin-1}
 \eea
Since $\vf$ is not present in the theory with action $S$, there is no corresponding equation. However, we still have the Noether identity  \eqref{Stueckelberg Noether} evaluated at $\vf =0$,
\bea
 -\pa^\m\frac{\d \widetilde{S}}{\d A^\m}\bigg|_{\vf=0}+m\,\frac{\d \widetilde{S}}{\d \vf}\bigg|_{\vf=0}=0~.
\eea
Thus, the equation of motion \eqref{massive_spin-1} implies 
\bea
 \frac{\d \widetilde{S}}{\d \vf}\bigg|_{\vf=0}=0~.
\eea
We conclude that the actions $\widetilde S$ and $S$ lead to equivalent dynamics. 

Our second example is Maxwell's theory 
\bea
S_{\rm Max} [A_\m]= - \frac 14 \int {\rm d}^4 x\, F^{\m \n} F_{\m\n }  ~, \qquad F_{\m\n} = \pa_\m A_\n - \pa_\n A_\m ~.
\eea
The corresponding Noether identity is 
\bea
    \pa_\m \frac{\d S_{\rm Max}}{\d A_\m}=  \pa_t\frac{\d S_{\rm Max}}{\d A_0}+  \pa_i\frac{\d S_{\rm Max}}{\d A_i}\equiv0~, 
 \label{Noether2}
 \eea
 and the equation of motion for the four-potential,  ${\d S_{\rm Max}}/{\d A_\m}= 0$, can be rewritten as 
 \bea
 \frac{\d S_{\rm Max}}{\d A_0}= - \grad \cdot \vec{E}=0~, \qquad \frac{\d S_{\rm Max}}{\d A_i}=0~.
 \label{EoM-Maxwell}
 \eea
 Let us choose the temporal gauge, $A_0=0$, and denote by $S[A_i]$ the resulting gauge-fixed action, 
 $S= S_{\rm Max}|_{A_0=0}$. The equation of motion for $S$,
\bea
\frac{\d S}{\d A_i}=\frac{\d S_{\rm Max}}{\d A_i}\bigg|_{A_0=0} =0~, 
\eea
coincides with the second equation in \eqref{EoM-Maxwell}. However, we have lost the Gauss law, 
$\grad \cdot \vec{E}=0$, since now the Noether identity \eqref{Noether2} implies only the relation $\pa_t \grad \cdot \vec{E}=0$. 
We see that the models $S_{\rm Max} $ and $S$ are not dynamically equivalent. 
The Gauss law must be added as a constraint when dealing with $S$.
 
 The difference between the two examples considered above is quite simple.
 In accordance with \eqref{Stueckelberg Gauge Transformations}, the gauge transformation acts on $\vf$ as a shift, $\d \vf = m \x$, while the transformation of $A_0$ involves a derivative, 
 $\d A_0 = \pa_t \x$. Coming back to the Pashnev theory, it follows from 
 \eqref{Other Independent Gauge Transformations} that the gauge variations of the fields 
 $| \psi^{(s-1)}\rangle $ and  $| \psi^{(s-2)}\rangle $ involve derivatives, and therefore a careful analysis is required to understand whether the gauge-fixed theory with action $S_{\rm alt}$ is equivalent to the original Pashnev theory.
 
Let $S_{\rm{P}}$ denote the action functional for the Pashnev theory, 
eq. \eqref{Action All Indpendent Fields}.
 From the gauge transformations \eqref{Other Independent Gauge Transformations} we derive the Noether identities 
\bsubeq\label{NI Pashnev}
\bea
    l_1\bigg|\frac{\d S_{\rm{P}}}{\d \j^{(s)}}\bigg\rangle +m\bigg|\frac{\d S_{\rm{P}}}{\d \j^{(s-1)}}\bigg\rangle-2 l_2^+l_1\bigg|\frac{\d S_{\rm{P}}}{\d \j^{(s-2)}}\bigg\rangle-6 m l_2^+\bigg|\frac{\d S_{\rm{P}}}{\d \j^{(s-3)}}\bigg\rangle &\equiv &0~,\\
    l_1\bigg|\frac{\d S_{\rm{P}}}{\d \j^{(s-1)}}\bigg\rangle+2 m\bigg|\frac{\d S_{\rm{P}}}{\d \j^{(s-2)}}\bigg\rangle-2 l_2^+l_1\bigg|\frac{\d S_{\rm{P}}}{\d \j^{(s-3)}}\bigg\rangle &\equiv &0~.
\eea
\esubeq
The equations of motion for this theory are:
\bea
\bigg| \frac{\d S_{\rm{P}}}{\d \j^{(s-i)}} \bigg\rangle =0~, \qquad i = 0,1,2,3~.
\eea
Imposing the gauge conditions \eqref{4.55}, which we will collectively label $G_{\rm{alt}}=0$, leads to the gauge-fixed action $S_{\rm{alt}}=S_{\rm{P}}|_{G_{\rm{alt}}=0}$, and the corresponding equations of motion are:
\bea
    \bigg|\frac{\d S_{\rm{alt}}}{\d\j^{(s)}}\bigg\rangle =\bigg|\frac{\d S_{\rm{P}}}{\d\j^{(s)}}\bigg\rangle\bigg|_{G_{\rm{alt}}=0}=0~,
    \qquad \bigg|\frac{\d S_{\rm{alt}}}{\d\j^{(s-3)}}\bigg\rangle =\bigg|\frac{\d S_{\rm{P}}}{\d\j^{(s-3)}}\bigg\rangle\bigg|_{G_{\rm{alt}}=0}=0~.
\eea
This leads to an important question, specifically: is there a way to recover the ``gauge-fixed'' 
equations of motion for the rank $s-1$ and $s-2$ fields? 
We still have the Noether identities \eqref{NI Pashnev} evaluated at $G_{\rm{alt}}=0$:
\bsubeq
\bea
    m\bigg|\frac{\d S_{\rm{P}}}{\d \j^{(s-1)}}\bigg\rangle\bigg|_{G_{\rm{alt}}=0}-2 l_2^+l_1\bigg|\frac{\d S_{\rm{P}}}{\d \j^{(s-2)}}\bigg\rangle\bigg|_{G_{\rm{alt}}=0}&=&0~,\\
    l_1\bigg|\frac{\d S_{\rm{P}}}{\d \j^{(s-1)}}\bigg\rangle\bigg|_{G_{\rm{alt}}=0}+2 m\bigg|\frac{\d S_{\rm{P}}}{\d \j^{(s-2)}}\bigg\rangle\bigg|_{G_{\rm{alt}}=0}&=&0~,
\eea
\esubeq
and they imply the following differential conditions:
\bsubeq
\bea
    (l_2^+(l_1)^2+m^2)\bigg|\frac{\d S_{\rm{P}}}{\d \j^{(s-1)}}\bigg\rangle\bigg|_{G_{\rm{alt}}=0}&=&0~,\\
    (l_1l_2^+l_1+m^2)\bigg|\frac{\d S_{\rm{P}}}{\d \j^{(s-2)}}\bigg\rangle\bigg|_{G_{\rm{alt}}=0}&=&0~.
\eea
\esubeq
These conditions do not lead to a recovery of the two equations of motion we have lost due to the presence of derivatives. Hence, we see that $S_{\rm{P}}$ and $S_{\rm{alt}}$ do not describe equivalent dynamics. Similar to the case of the Maxwell action, the lost equations of motion must be added as constraints to the action.

We now wish to carry out a similar analysis for our gauge conditions \eqref{gauge_condition}, for which it is appropriate to work with the action $S_{\rm{KZ}}$ given in \eqref{Full Action for spin s}. 
The corresponding Lagrangian in the oscillator formalism reads
\bea
\label{KZ Lagrangian in Oscillator Form}
    \mathcal{L}^{(s)} &=& \sum^{s}_{k=0}\mathcal{L}_c(\f^{(s-k)})\non \\
    \non \\
    &=& \sum^{s}_{k=0}\frac{(s-k)!}{2}\bigg[\langle\f^{(s-k)}|\bigg(-l_0+l_1^+l_1+2l_2^+l_0 l_2-(l_1^+)^2l_2-l_2^+(l_1)^2+l_2^+l_1^+l_1l_2\non \\
    &&+2d_{s-k}+8\frac{e_{s-k}}{(s-k)(s-k-1)}l_2^+l_2\bigg)|\f^{(s-k)}\rangle\non\\
    &&+ \bigg(\ri \frac{a_{s-k}}{s-k}\langle\phi^{(s-k-1)}|l_1|\phi^{(s-k)}\rangle+2\ri \frac{b_{s-k}}{(s-k)(s-k-1)}\langle\phi^{(s-k)}|l_2^+l_1|\phi^{(s-k-1)}\rangle\non \\
    && +4\ri \frac{c_{s-k}}{(s-k)(s-k-1)(s-k-2)}\langle\f^{(s-k-1)}|l_2^+l_1l_2|\phi^{(s-k)}\rangle\non \\
    &&-2\frac{f_{s-k}}{(s-k)(s-k-1)}\langle\phi^{(s-k)}|l_2^+|\phi^{(s-k-2)}\rangle+\rh.\rc.\bigg)\bigg]~.
\eea
We recall that double traceless fields can be decomposed into two traceless fields via
\bea
\label{KZ field general}
    |\f^{(s-k)}\rangle &=& |\o^{(s-k)}\rangle+\frac{1}{d+2s-2k-4}l_{2}^+|\vf^{(s-k-2)}\rangle~.
\eea
From \eqref{Zinoviev Gauge Transformation Oscillator Form} and \eqref{KZ field general},  the gauge transformation laws of the traceless fields prove to be
\bsubeq \label{Traceless Gauge Transformation KZ}
\bea
    \d|\o^{(s-k)}\rangle &=& \a_{s-k}|\x^{(s-k)}\rangle-\frac{\ri}{s-k}\left(\frac{2}{d+2s-2k-4}l_2^+l_1-l_1^+\right)|\x^{(s-k-1)}\rangle~, \\
    \non\\
    \d|\vf^{(s-k)}\rangle &=& -(d+2s-2k)\b_{s-k+2}|\x^{(s-k)}\rangle+\frac{2\ri}{s-k+2}l_1|\x^{(s-k+1)}\rangle~.
\eea
\esubeq
The corresponding Noether identities are
\bea \label{KZ NI}
 &&   \frac{\ri}{s-k+1} \bigg\{
    l_1\bigg|\frac{\d S_{\rm{KZ}}}{\d\o^{(s-k+1)}}\bigg\rangle+
    2
    \bigg(l_1^+-\frac{2}{d+2s-2k-4}l_2^+l_1\bigg)\bigg|\frac{\d S_{\rm{KZ}}}{\d\vf^{(s-k-1)}}\bigg\rangle
\bigg\}    \non \\
 &&  \quad  -\a_{s-k}\bigg|\frac{\d S_{\rm{KZ}}}{\d\o^{(s-k)}}\bigg\rangle+(d+2s-2k)\b_{s-k+2}\bigg|\frac{\d S_{\rm{KZ}}}{\d\vf^{(s-k)}}\bigg\rangle\equiv0~, \qquad 1\leq k\leq s~.
\eea

We now impose the gauge conditions \eqref{gauge_condition}, which we collectively denote $G_{\rm{SH}}=0$. From \eqref{Traceless Gauge Transformation KZ}, we see that the gauge parameters are constrained by the equations
\bsubeq
\bea
    \a_{s-k}|\x^{(s-k)}\rangle&=&\frac{\ri}{s-k}\left(\frac{2}{d+2s-2k-4}l_2^+l_1-l_1^+\right)|\x^{(s-k-1)}\rangle~, \qquad 1\leq k \leq s-1~,~~ \\
    |\x^{(0)}\rangle &=&0~,
\eea
\esubeq
which imply
\bea
    |\x^{(s-k)}\rangle =0~,\qquad 1\leq k \leq s~.
\eea
Therefore, the gauge conditions \eqref{gauge_condition} completely fix the gauge freedom. 

The gauge-fixed action $S_{\rm{SH}}=S_{\rm{KZ}}|_{G_{\rm{SH}}=0}$ leads to the following equations of motion:
\bsubeq
\bea
   \bigg|
   \frac{\d S_{\rm{SH}}}{\d\o^{(s)}}
   \bigg\rangle 
   = 
   \bigg|
   \frac{\d S_{\rm{KZ}}}{\d\o^{(s)}}\bigg\rangle\bigg|_{G_{\rm{SH}}=0}&=&0 ~,\\
    \non \\
    \bigg|
    \frac{\d S_{\rm{SH}}}{\d\vf^{(s-k)}}
    \bigg\rangle 
    = 
    \bigg|
    \frac{\d S_{\rm{KZ}}}{\d\vf^{(s-k)}}
    \bigg\rangle\bigg|_{G_{\rm{SH}}=0}&=&0~,\qquad 2\leq k \leq s~.
\eea
\esubeq
Assuming these equations hold, the Noether identities \eqref{KZ NI} evaluated at $G_{\rm{SH}}=0$ are
\bsubeq
\bea
    \frac{\ri}{s-k+1}l_1\bigg|\frac{\d S_{\rm{KZ}}}{\d\o^{(s-k+1)}}\bigg\rangle\bigg|_{G_{\rm{SH}}=0}-\a_{s-k}\bigg|\frac{\d S_{\rm{KZ}}}{\d\o^{(s-k)}}\bigg\rangle\bigg|_{G_{\rm{SH}}=0}&=&0~,\qquad 2\leq k \leq s~,\\
    \non \\
    \bigg|\frac{\d S_{\rm{KZ}}}{\d\o^{(s-1)}}\bigg\rangle\bigg|_{G_{\rm{SH}}=0}&=&0~,
\eea
\esubeq
which imply 
\bea
    \bigg|
    \frac{\d S_{\rm{KZ}}}{\d\o^{(s-k)}}
    \bigg\rangle\bigg|_{G_{\rm{SH}}=0}&=&0~,\qquad 1\leq k \leq s~.
\eea
We conclude that $S_{\rm{KZ}}$ and $S_{\rm{SH}}$ lead to equivalent dynamics.

We can consider each of the two sets of gauge conditions, \eqref{gauge_condition} and \eqref{4.55},
as specific cases of the more general class of gauge conditions\footnote{To realise \eqref{4.55} in such a form, we make use of the equivalence analysed in section \ref{Pashnev-KZ}.}
\bsubeq\label{Linear Gauge Condition}
\bea
    A_k|\vf^{(s-k)}\rangle-B_k|\o^{(s-k)}\rangle&=&0~, \qquad B_k\neq0~, \qquad 
    2\leq k\leq s~,\\
    \non\\
    |\o^{(s-1)}\rangle&=&0~,\qquad\qquad\qquad\qquad\qquad
\eea
\esubeq
where $(A_k,B_k)$ is a constant non-zero real two-vector defined modulo rescalings
\bea
(A_k,B_k)\sim\l(A_k,B_k)~, \qquad\l\neq 0~.
\eea
By repeating the above analysis for the gauge conditions \eqref{Linear Gauge Condition},
 we find that \eqref{gauge_condition} is the unique choice within this class of gauge conditions with the crucial properties: (i) it completely fixes the gauge freedom; and (ii) the dynamics generated by 
the corresponding gauge fixed action is equivalent to that described by the gauge-invariant action
 \eqref{Full Action for spin s}. 
It follows that the $d$-dimensional generalisation to the Singh-Hagen model, which we derived in section \ref{Singh Hagen Correspondence}, is unique.

\end{document}